\documentclass[aps,pra,reprint,floatfix,superscriptaddress,nofootinbib,amsmath,amssymb,longbibliography]{revtex4-2}
\pdfoutput=1
\usepackage[utf8]{inputenc}
\usepackage[T1]{fontenc}
\usepackage{lmodern,microtype}
\usepackage{ragged2e}
\usepackage{tikz}
\usetikzlibrary{arrows.meta,positioning}
\usepackage[colorlinks=true,linkcolor=blue,allcolors=blue]{hyperref}
\hypersetup{bookmarksdepth=2,pdftitle={On the Completeness of Quantum Mechanics: Sources, Records and Probabilities},pdfauthor={Mustafa Bakr}}
\begin{document}
\title{On the Completeness of Quantum Mechanics: Sources, Records and Probabilities}
\author{Mustafa Bakr}
\email{mustafa.bakr@physics.ox.ac.uk}
\affiliation{Department of Physics, Clarendon Laboratory, University of Oxford, United Kingdom}
\affiliation{Centre for Quantum Technologies, National University of Singapore, Singapore}
\begin{abstract}
Can quantum mechanics be complete without being fundamental? Under stated source, detector, and statistical conditions, we reconstruct the quantum representation and Born probabilities and prove complete algebraic closure of the admitted quantum predictions without requiring a collapse postulate. We further show that many familiar quantum paradoxes arise from conflating coherent processes with retained records or probabilities with realized outcomes. We also ask whether mathematics as a whole, or the source architecture itself, emerges from a deeper structure.
\end{abstract}
\maketitle
\section{Introduction}
\label{sec:II-introduction}
\label{sec:I-introduction-physical}
The question of whether quantum mechanics is complete is older than many of its applications. Einstein, Podolsky and Rosen asked whether the quantum description contains a counterpart for every element of physical reality~\cite{EPR}. Englert treats the event as a preexisting concept of quantum theory and Born's rule as the link between the formalism and event probabilities~\cite{Englert2013}. A disagreement about completeness can therefore begin before any calculation in what we ask a physical theory to explain.

Consider an interference experiment. An individual run leaves a mark on a detector. Many runs produce a pattern, and Born's rule connects that pattern to the amplitudes calculated from the quantum state~\cite{Born1926}. The interaction with the detector, the probabilities of its possible readings and the occurrence of one reading are related questions. They are nevertheless different questions. A calculation of a correlation between a system and a detector does not, by itself, explain why a particular reading occurs. Schr\"odinger's cat brings this distinction into familiar terms~\cite{Schrodinger1935}. A theory can determine every prediction in its domain without being the most fundamental description of the processes involved. Hence, completeness of prediction and fundamentality are different questions. This distinction leads us to the physical laws from which a quantum description might be reconstructed.

Take different field theories as an example. They describe different interactions and yet quantum physics uses a common mathematical language: complex amplitudes, Hilbert spaces, operators and probabilities. We ask which features of this language reflect a general structure of physical processes and which depend on the particular dynamical law. Could this common structure be reconstructed from a deeper physical description? If so, what would completeness then mean? We therefore ask both what the quantum description predicts and what its underlying source must supply. In~~\cite{FPTScalar,FPTQED}, the finite-proper-time scalar field theory and spinor QED provide concrete examples of specified source laws. Here the field laws are inputs to the reconstruction. Their common quantum representation is the proposed consequence. The reconstruction does not choose those laws or their couplings.

Throughout the paper, we use the word \emph{source} for a physical law together with the conditions under which its processes take place. These include the physical domain, boundaries, finite resolution and rules for joining processes. In an experiment, they would also include the preparation, apparatus and controls. This is the proposed starting point. The Hilbert space is a representation to be constructed from it. Calling that representation emergent leaves us a task: establishing the properties of the source that make the construction possible.

For measurement terminology we use standard modern operator language while making the connection to Englert explicit. A density operator is the statistical operator describing a preparation. An operator $0\le E\le I$ associated with one detector alternative is called an \emph{effect}; Englert calls the same object a \emph{probability operator}, and calls a normalized family of such operators a probability-operator measurement (POM), commonly termed a POVM. We retain the word \emph{effect} until the statistical theorem assigns probabilities. By a \emph{reader} we mean the terminal measurement apparatus or readout for the declared record. The closest analogue of Englert's \emph{document} is our retained record, but we distinguish its physical production (registration), its persistence under the declared continuation (retention), and the realization of one record in a completed run. Englert's \emph{state reduction} is the conditional updating used below after a record is known; it is not an additional collapse dynamics.

There is a further question even when that construction succeeds. Why should the squared length of a branch give the chance of a detector record? Geometry alone does not supply an interpretation as probability. The physical and statistical conditions entering that interpretation must therefore be stated and their consequences derived. The same distinction applies to collapse. Updating a prediction after a record is known and proposing a physical process that produces one outcome answer different questions.

\begin{figure*}[t]
\centering
\begin{tikzpicture}[>=Latex,font=\small,
  box/.style={draw,rounded corners=1.5pt,align=center,text width=3.8cm,minimum height=1.0cm,inner sep=5pt}]
\node[box] (src) at (0,0) {Source law and physical domain\\$Z,\mathcal D,\mathsf{Res}$};
\node[box] (rep) at (5.7,0) {Quantum representation\\$\mathcal H,\ T_M$};
\node[box] (rec) at (11.4,0) {Detector record effects\\(Englert: probability operators)\\$E_r=V^\dagger Q_rV$};
\draw[->] (src) -- node[above,align=center,font=\scriptsize,text width=1.50cm] {positive\\pairing and\\composition} (rep);
\draw[->] (rep) -- node[above,align=center,font=\scriptsize,text width=1.50cm] {physical\\writing and\\retention} (rec);
\node[box] (axioms) at (0,-2.2) {Statistical conditions\\preparation and mixing\\readout calibration};
\node[box] (born) at (5.7,-2.2) {Born probabilities\\$p_r=\operatorname{Tr}(\rho E_r)$};
\node[box] (hist) at (11.4,-2.2) {Complete recorded histories\\joint and conditional laws};
\draw[->] (axioms) -- (born);
\draw[->] (rec.south) -- ++(0,-.35) -| (born.north);
\draw[->] (born) -- node[above,align=center,font=\scriptsize,text width=1.50cm] {admitted\\complete\\instruments} (hist);
\end{tikzpicture}
\caption{The physical questions in the reconstruction. Each arrow requires the conditions stated in the text. The upper row constructs represented objects without assigning chances to records. The displayed effect uses an isometric apparatus. A complete nonunitary process uses its dual map. The statistical conditions enter separately. The resulting history law specifies probabilities. It does not select the record realized in one run.}
\label{fig:source-record-probability}
\end{figure*}
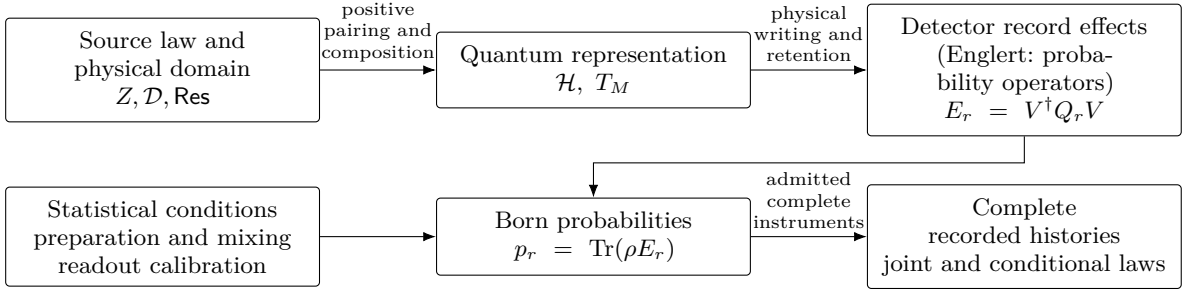

We examine these questions in that order, from the source to the detector and then to its probabilities. The aim is to identify what a source must supply for the quantum description and its probability rule to follow, and what remains a question about the source's own dynamics. Completeness here concerns predictions for the experiments covered by the stated conditions. Whether those predictions also give a complete account of physical reality is the question with which the discussion began. Figure~\ref{fig:source-record-probability} separates the source and detector construction from the statistical step. The short preparation argument in Sec.~\ref{sec:I-born-main} is the principal Born derivation. Section~\ref{sec:causality-route} examines an alternative set of physical assumptions. The appendices retain the detailed tests and counterexamples.

\section{Primitive sources and quantum representation}
\label{joint:source}
\subsection{The primitive source and its physical domain}
\label{sec:II-enhanced-primitive-source}
What information is required before a physical process can be calculated? Start with allowed physical input and output boundaries \(\Sigma\), and processes that connect them.  Compatible processes can be joined in sequence. Independent processes can be placed alongside one another.  There is an identity process at each boundary and a specified reversal operation.  Write $\mathsf{Src}$ for these boundaries, processes and joining rules.\footnote{Their formal mathematical name is a dagger-monoidal category. Objects are boundaries, arrows are allowed processes, the product places independent processes together, and the dagger is the specified reversal. The compatibility rules, not the name, are the hypotheses used here.} The symbol $I$ denotes the empty boundary, with no external input or output. It does not denote a prepared vacuum state.  Thus
\begin{equation}
M:\Sigma_i\longrightarrow\Sigma_f
\end{equation}
are finite physical source regions or source processes.  We write \(\circ\) for joining processes in sequence, \(\otimes\) for placing independent processes alongside one another, and \(M\mapsto M^\dagger\) for the specified reversal.  Reversal has not been assumed to be an operation that physically undoes the process.  We reserve \emph{realized event} for a definite record in a completed run. A primitive process \(M\) is not such an occurrence.  A source preparation is a process
\begin{equation}
s:I\longrightarrow\Sigma.
\label{eq:II-source-preparation-type}
\end{equation}
For each finite experiment \(\mathfrak E\), \(\mathcal D_{\mathfrak E}\) denotes its physical domain data: bulk region, boundary conditions, regularity/form domains, and any bundle, spin, gauge, constraint or relational data needed to specify the realization.  Gauge-equivalent descriptions represent the same physical configuration. Their identification and any constraints belong to the specified source. A single global gauge choice is not assumed.

The primitive datum is written schematically as
\begin{equation}
\boxed{
\mathbf K_{\mathfrak E}^{\sharp}
=
\bigl(
\mathsf{Src},\circ,\otimes,\dagger,
\mathcal D_{\mathfrak E},
\mathsf{Res},
\mathcal G,
Z
\bigr).}
\label{eq:II-Ksharp}
\end{equation}
Here $\mathsf{Res}$ specifies physical resolution and $\mathcal G$ the symmetry, gauge and constraint structure. The law $Z$ supplies the sector dynamics. Each field theory supplies a realization of this datum, including its variables, constraints and boundaries. The whole source specification is distinct from an external current $J$ used to differentiate a generating functional. Changing its domain or physical resolution can change its predictions.

For example, a finite-proper-time construction uses the spectral function
\[
 R_{s_0}(P)=\int_{s_0}^{\infty}e^{-sP}\,ds
          =e^{-s_0P}P^{-1}
\]
for a specified self-adjoint operator $P\ge p_{\min}I$ with $p_{\min}>0$. The operator, its domain and the endpoint together determine this function. The scalar field and spinor-QED constructions supply spectral and interaction rules~\cite{FPTScalar,FPTQED}. The question here is which additional properties of such a law establish a quantum representation and connect it to detector records. Collecting the data in a tuple does not establish any of those properties.

The source functional \(Z\) is defined on the closed source processes generated by composition and dagger and takes values in the scalar endomorphism system
\begin{equation}
\mathbb S:=\operatorname{End}(I).
\end{equation}
The primitive datum specifies possible interactions and boundaries. Identifying a boundary as carrying a readable record requires the registration construction below.

Why Nature realizes a particular $Z$ and its parameters remains a question for that source theory.
\subsubsection*{Doubled source pairing before scalar realization}
For source preparations \(s,t:I\to\Sigma\), write \(Z_\Sigma\) for the source functional restricted to closed composites through the boundary \(\Sigma\), and define the \(\mathbb S\)-valued doubled pairing
\begin{equation}
\kappa_\Sigma(s,t)
=
Z_\Sigma(s^\dagger\circ t).
\label{eq:II-Kernel-from-source}
\end{equation}
No complex coefficients or probabilities enter this pairing. After identifying the scalar system, the physical kernel must satisfy the positivity condition below.

\subsection{Physical resolution and compatible source families}
\label{sec:II-perfect-resolution}
The finite-proper-time source examples carry a common set of physical-resolution data together with sector-specific dynamical parameters~\cite{FPTScalar,FPTQED}.  Let \(\mathcal D\) denote the global physical domain, \(\mathfrak B\) its boundary realization, \(s_0>0\) the retained lower resolution of a complete virtual/source history, \(t_i\) and \(t_f\) the initial and final boundary times, \(\mathcal R\) the retained detector/channel resolution, and \(\boldsymbol\lambda_{\rm sector}\) the remaining sector-dependent dynamical data.  Define the finite duration by \(\tau:=t_f-t_i\). It is the duration of the specified experiment, not a new universal minimum time. Collect the data as
\begin{equation}
\boxed{
\begin{aligned}
\mathfrak E^\sharp
&=(\mathcal D,\mathfrak B,s_0,[t_i,t_f],\mathcal R;\boldsymbol\lambda_{\rm sector}),\\
s_0&>0,\qquad \tau<\infty .
\end{aligned}}
\label{eq:II-Esharp}
\end{equation}
Increasing a numerical basis at fixed $\mathfrak E^\sharp$ improves the description of the same experiment. Changing $s_0$, $\tau$, $\mathcal R$ or the physical domain changes the source problem. An embedding $j$ at fixed physical resolution preserves the reconstructed pairings precisely when
\[
 K'(js,jt)=K(s,t).
\]
It then induces an isometry. In the scalar and spinor constructions, functional insertions divide an existing history, while interaction sewing joins histories~\cite{FPTScalar,FPTQED}. Their cutting and deletion results establish compatibility with independently constructed daughter processes on the stated continuation domains. The complete-history endpoint prescription is shared, while field content, statistics and gauge constraints remain sector specific. These amplitude statements do not assign detector probabilities.

\label{sec:II-primitive-emergent}
Here \emph{emergent} means constructed from source data without being used to define those data. It does not mean that the source law or mathematics itself has been derived.

\subsection{From the source pairing to the quantum representation}
\label{sec:II-universal-reconstruction}
Which parts of the familiar formalism follow from the source? We first
identify its scalar values, then the inner product and state space,
and finally the operators representing its processes. The mathematical
constructions are standard once their hypotheses hold. The physical
question is whether one specified source supplies those hypotheses
without using the representation to define them.
\subsubsection*{Scalars and the complex realization}
A process with the empty boundary as both input and output is called a source scalar.  The compatibility of sequential and independent composition, called interchange, gives
\begin{equation}
(a\otimes1)(1\otimes b)=(1\otimes b)(a\otimes1),
\end{equation}
and hence
\begin{equation}
\boxed{ab=ba.}
\end{equation}
Thus the compositional structure supplies a commutative scalar system \(\mathbb S=\operatorname{End}(I)\), with involution \(a^*:=a^\dagger\) inherited from the source dagger.  Interchange alone does not make this system a field, nor does it select \(\mathbb C\).

\paragraph{A continuous-phase scalar-selection theorem.}
Interchange proves commutativity, but a further physical/algebraic bridge is required before the scalar system can be identified with \(\mathbb C\).  One sufficient route makes the role of coherent phase explicit.  Suppose the source scalar system is supplied with addition and real scalar multiplication, compatible with composition, making it a finite-dimensional associative unital real \(*\)-algebra in which every nonzero scalar is reversible. This additive algebra structure is an additional hypothesis. Joining and reversal alone do not supply it.  Let \(\mathbb S^\times\) denote the group of invertible scalars.  Suppose also that the source admits a nonconstant continuous one-parameter family of reversible scalar phases
\begin{equation}
\begin{aligned}
 u&:\mathbb R\longrightarrow\mathbb S^\times,
 &u(0)&=1,\\
 u(\theta+\phi)&=u(\theta)u(\phi),
 &u(\theta)^*u(\theta)&=1 .
\end{aligned}
 \label{eq:II-continuous-source-phase}
\end{equation}
Then the commutative scalar algebra is \(*\)-isomorphic to \((\mathbb C,\text{complex conjugation})\).

Appendix~\ref{app:representation-details} gives the argument. Commutativity and reversibility restrict this finite-dimensional real algebra to $\mathbb R$ or $\mathbb C$. A continuous nontrivial phase excludes $\mathbb R$ and fixes conjugation as the involution. These conclusions hold within the stated algebra class.

Under the stated scalar-selection hypotheses, fix a faithful unital $*$-compatible scalar identification
\begin{equation}
\chi:\mathbb S\overset{\sim}{\longrightarrow}\mathbb C
\end{equation}
for the complex realization used in the remainder of the paper.  This is a reconstruction hypothesis within the stated candidate class, not a theorem that the primitive joining and reversal rules alone classify every possible scalar system.

The represented doubled source kernel is then
\begin{equation}
K_\Sigma(s,t):=\chi\!\left(\kappa_\Sigma(s,t)\right).
\label{eq:II-complex-source-kernel}
\end{equation}
The Hilbert reconstruction requires, as an independent source condition, that for every finite family of source preparations and complex coefficients,
\begin{equation}
\boxed{
\sum_{i,j}\bar c_i c_j K_\Sigma(s_i,s_j)\ge0.}
\label{eq:II-source-positive-double}
\end{equation}
Positivity is required for every finite pairing matrix, not just its diagonal or two-preparation submatrices. It is a physical source condition unless proved in the specified sector. The following construction supplies its Hilbert representation, without assigning record probabilities.

One possible source-specific proof of positivity would establish
\begin{equation}
 K_\Sigma(s,t)=\int \overline{A_Z(X|s)}A_Z(X|t)\,d\nu_Z(X),
 \label{eq:physical-fold-identification}
\end{equation}
where $\nu_Z$ is a positive physical channel measure and the amplitudes are square integrable. Finite linear combinations then give an integral of an absolute square. The essential step is equality with the pairing already defined by $Z(s^\dagger\circ t)$. Defining a new Gram kernel by the right-hand side proves that kernel positive, but does not prove the equality for the original source. Nor does a positive channel measure by itself identify the chances of detector records.

The conclusion \(\mathbb S\simeq\mathbb C\) under the stated sufficient conditions selects the scalar field of the standard complex representation.  It does not locate a universal ``first quantum object''.  Complex numbers are not sufficient for nonclassicality, since an ordinary classical stochastic theory embeds in complex Hilbert space using a diagonal commuting operator algebra.  They are not necessary either: the real two-qubit vector \(|\Phi^+\rangle=(|00\rangle+|11\rangle)/\sqrt2\) together with the Pauli matrices \(X\) and \(Z\) and the entirely real observables \(A_0=Z\), \(A_1=X\), \(B_0=(Z+X)/\sqrt2\), \(B_1=(Z-X)/\sqrt2\) gives the CHSH value \(2\sqrt2\).  This example excludes Bell-local descriptions with setting-independent preparations. It does not rule out every classical representation or hidden-variable dynamics. Neither complex scalars nor noncommutativity alone are sufficient criteria for this distinction.

\subsubsection*{Hilbert-space reconstruction and minimal uniqueness}
Let \(\mathcal V_\Sigma\) be the complex algebraic span of formal source vectors \(|s\rangle\).  These formal sums provide a linear representation.  Their inclusion does not establish that every superposition can be physically prepared. That requires the source and control specification.  Define
\begin{equation}
\begin{gathered}
\langle s|t\rangle=K_\Sigma(s,t),\\
\mathcal N_\Sigma=\{v\in\mathcal V_\Sigma:\langle v|v\rangle=0\}.
\end{gathered}
\end{equation}
Then
\begin{equation}
\boxed{
\mathcal H_\Sigma
=
\overline{\mathcal V_\Sigma/\mathcal N_\Sigma}.}
\label{eq:II-H-source-complete}
\end{equation}
For a small example, suppose two preparations have pairing matrix $\left(\begin{smallmatrix}1&1\\1&1\end{smallmatrix}\right)$.  The formal vector $a|s_1\rangle+b|s_2\rangle$ has squared length $|a+b|^2$.  The difference $|s_1\rangle-|s_2\rangle$ has zero length, so identifying zero-length differences leaves one complex direction.  This illustrates the construction from a supplied positive pairing. It gives no detector probability by itself.

Cauchy--Schwarz, obtained from positivity of $\langle v+zw,v+zw\rangle$ for every complex $z$, makes each null vector orthogonal to every vector. The quotient is therefore well defined. Completion supplies its norm limits.

If two realizations reproduce all source pairings and the represented source vectors have dense span, the source-state correspondence extends uniquely to a unitary.  This is uniqueness up to unitary equivalence of the minimal representation.  Adding an inaccessible orthogonal summand gives a nonminimal counterexample and is not identified with a new physical realization.

\subsubsection*{Transport of the source pairing}
\label{sec:source-transport}
Conservation can be tested in source coordinates before choosing an orthonormal basis. On a finite source span let $G(t)$ be the independently supplied Hermitian pairing matrix and let the invertible evolution matrix $B(t)$ obey
\[
 \dot B(t)=L_Z(t)B(t),\qquad B(0)=I.
\]
Assume $G$ is continuously differentiable and $L_Z$ is continuous on the time interval. Define the source discrepancy
\begin{equation}
 D_Z(t)=\dot G(t)+L_Z(t)^\dagger G(t)+G(t)L_Z(t).
 \label{eq:source-discrepancy}
\end{equation}
The product rule gives the exact identity
\begin{equation}
\begin{split}
 B(t)^\dagger G(t)B(t)-G(0)\\
 =\int_0^t B(u)^\dagger D_Z(u)B(u)\,du.
\end{split}
\label{eq:source-transport}
\end{equation}
If $G(0)\ge0$ and $D_Z=0$, the pairing stays positive and $B$ transports its null space exactly. It consequently induces an onto isometry on the positive quotients. This conclusion uses the same pairing and dynamics supplied by the source. Defining $G(t)=B(t)^{-\dagger}G(0)B(t)^{-1}$ would make conservation true by construction and would not prove it for an independently specified physical pairing.

The identity also gives an error test. If $G(0)\ge mI$ with $m>0$ and
\[
 d(t)=\int_0^t\|B(u)\|^2\|D_Z(u)\|\,du<m,
\]
then
\[
 G(t)\ge\frac{m-d(t)}{\|B(t)\|^2}I.
\]
A lower bound on $G(0)$ is essential for this conclusion. With $B=I$ and $G(t)=\operatorname{diag}(1,-\epsilon t)$, the discrepancy can be arbitrarily small, while positivity fails for every $t>0$.

When $G(t)>0$, set $S(t)=G(t)^{1/2}$. Orthonormal coordinates give
\[
 U(t)=S(t)B(t)S(0)^{-1}.
\]
If $D_Z=0$, this is unitary and its generator is
\begin{equation}
 H_Z(t)=i\hbar\bigl[\dot S(t)S(t)^{-1}
                  +S(t)L_Z(t)S(t)^{-1}\bigr].
 \label{eq:source-moving-generator}
\end{equation}
Differentiating $U^\dagger U=I$ proves $H_Z^\dagger=H_Z$. The term containing $\dot S$ is required when the pairing coordinates change with time. These are finite-span statements. An unbounded or infinite-dimensional source requires its own domains and convergence estimates.

\subsubsection*{Disjoint source composition and the tensor-product state space}
For physically admitted independent boundaries, assume a bilinear preparation product
\begin{equation}
 \boxtimes:\mathcal V_{\Sigma_A}\times\mathcal V_{\Sigma_B}
 \longrightarrow\mathcal V_{\Sigma_A\otimes\Sigma_B}
 \label{eq:II-source-product-preparations}
\end{equation}
and factorization of the source pairing,
\begin{equation}
\begin{split}
 K_{\Sigma_A\otimes\Sigma_B}(s_A\boxtimes s_B,t_A\boxtimes t_B)\\
 =K_{\Sigma_A}(s_A,t_A)K_{\Sigma_B}(s_B,t_B).
\end{split}
\label{eq:II-source-kernel-factorization}
\end{equation}
Multiplicativity of $Z$ and $\chi$ on disjoint closed processes is sufficient for this identity. If the product preparations also span a dense subspace, their isometry extends onto the composite space,
\begin{equation}
 \mathcal H_{\Sigma_A\otimes\Sigma_B}
 \simeq\mathcal H_{\Sigma_A}\otimes\mathcal H_{\Sigma_B}.
 \label{eq:II-source-tensor-product-theorem}
\end{equation}
Appendix~\ref{app:representation-details} gives the proof. Source associativity and the empty boundary induce the corresponding Hilbert-space identifications. Gauge or gravitational constraints can obstruct factorization or density for a proposed subsystem split. Such a split requires its constrained boundary algebra, rather than an assumed unrestricted tensor product.

\subsubsection*{Representation of source processes}
Let \(M:\Sigma\to\Sigma'\) be an admissible source process and extend its action linearly to the algebraic source span \(\mathcal V_\Sigma\).  Suppose there is a finite constant \(c_M\ge0\) such that
\begin{equation}
K_{\Sigma'}(Mv,Mv)
\le c_M K_\Sigma(v,v)
\qquad(v\in\mathcal V_\Sigma).
\label{eq:II-source-process-bound}
\end{equation}
Then a null vector for \(K_\Sigma\) is mapped to a null vector for \(K_{\Sigma'}\), so
\begin{equation}
\boxed{
T_M[v]=[Mv]=[M\circ v]}
\label{eq:II-source-process-representation}
\end{equation}
is representative independent on the quotient.  Moreover
\begin{equation}
\|T_M[v]\|^2
\le c_M\|[v]\|^2,
\end{equation}
so \(T_M\) extends uniquely to a bounded operator \(\mathcal H_\Sigma\to\mathcal H_{\Sigma'}\) with \(\|T_M\|\le\sqrt{c_M}\).  Source composition gives
\begin{equation}
T_{\mathrm{id}}=I,
\qquad
T_{N\circ M}=T_NT_M .
\label{eq:II-source-functoriality}
\end{equation}
Suppose the dagger is compatible with the source pairing. For $s:I\to\Sigma$ and $t:I\to\Sigma'$,
\begin{equation}
K_{\Sigma'}(Ms,t)
=
K_\Sigma(s,M^\dagger t),
\end{equation}
and density of the represented source spans gives
\begin{equation}
\boxed{T_{M^\dagger}=T_M^\dagger.}
\label{eq:II-source-dagger-representation}
\end{equation}
Unbounded generators require invariant domains or a separate quadratic-form construction. The bounded estimate alone does not establish those extensions.

\subsubsection*{Symmetry and generators}
An invertible source symmetry with $g^\dagger=g^{-1}$ and $K_\Sigma(gs,gt)=K_\Sigma(s,t)$ gives a unitary $U(g)=T_g$. Strong continuity of a one-parameter action supplies a self-adjoint generator by Stone's theorem. Starting from an infinitesimal Lie algebra additionally requires a common invariant domain, controlled commutators, integrability and compatibility with global topology and any central extension. A physical symmetry preserves the boundary and operator domain.

\subsubsection*{Noncommutativity, joint information and the uncertainty relations}
Commuting scalar amplitudes do not imply commuting source operations. Their represented order can matter. For self-adjoint $A,B$ and a normalized $\psi\in\operatorname{Dom}(A)\cap\operatorname{Dom}(B)$, write $A_0=A-\langle A\rangle_\psi I$ and similarly for $B_0$. Cauchy--Schwarz gives
\begin{equation}
 (\Delta_\psi A)^2(\Delta_\psi B)^2
 \ge |\langle A_0\psi,B_0\psi\rangle|^2.
 \label{eq:II-weak-uncertainty}
\end{equation}
If also $\psi\in\operatorname{Dom}(AB)\cap\operatorname{Dom}(BA)$, this is
\begin{equation}
\begin{split}
 (\Delta_\psi A)^2(\Delta_\psi B)^2
 \ge\tfrac14|\langle\{A_0,B_0\}\rangle_\psi|^2\\
 {}+\tfrac14|\langle[A,B]\rangle_\psi|^2,
\end{split}
\label{eq:II-Schrodinger-Robertson}
\end{equation}
and implies Robertson's relation~\cite{Robertson1929},
\begin{equation}
 \Delta_\psi A\,\Delta_\psi B
 \ge\tfrac12|\langle[A,B]\rangle_\psi|.
 \label{eq:II-Robertson}
\end{equation}
These are Hilbert-space spreads. They become statistical preparation spreads when the corresponding measurements satisfy the probability theorem below. An error--disturbance statement requires its own apparatus model.

For a source symmetry represented by a Weyl canonical pair, differentiation on the common invariant core gives $[X,P]=i\hbar I$ and hence $\Delta X\Delta P\ge\hbar/2$. A finite-dimensional truncation cannot obey that exact commutator, whose trace would be nonzero. The physical parameters $s_0$ and $\mathcal R$ specify admitted histories and record distinctions. They do not replace these uncertainty relations.

A product bound alone also leaves other uncertainty questions open~\cite{EnglertUncertainty2023}. For Pauli observables $X,Z$ and a Bloch vector $s$ with $\|s\|\le1$,
\[
 (\Delta X)^2+(\Delta Z)^2=2-s_x^2-s_z^2\ge1,
\]
even when the commutator expectation vanishes. The full uncertainty boundary is a separate optimization problem.

Strongly commuting self-adjoint observables have a joint spectral measure, and consequently a joint probability distribution after the statistical step. That distribution can have nonzero spreads. Noncommuting spectral projections admit no common sharp measure with all their spectral marginals. The classical record algebra below concerns distinctions made jointly retainable by the apparatus.

\subsubsection*{Interventions and physical process maps}
For declared von Neumann input and output algebras, a represented linear intervention $\mathcal E$ is completely positive if $I_n\otimes\mathcal E$ is positive for every finite matrix auxiliary system. Normality means preservation of bounded increasing limits. Both properties, and extension to the full declared algebra when initially defined on a smaller set, require source justification.

Normalization is independent: $\mathcal E(I_{\rm out})\le I_{\rm in}$, with equality for a complete deterministic channel. The normal CP map $\mathcal E(A)=2A$ is a counterexample without normalization. Stinespring's theorem gives~\cite{Stinespring1955}
\begin{equation}
 \mathcal E(A)=W_{\mathcal E}^\dagger\pi_{\mathcal E}(A)W_{\mathcal E},
 \qquad W_{\mathcal E}^\dagger W_{\mathcal E}\le I,
 \label{eq:II-Stinespring-complete}
\end{equation}
where $\pi_{\mathcal E}$ is a representation and $W_{\mathcal E}$ a dilation operator, isometric for a unital channel. This mathematical dilation is distinct from a physically complete conservative process. Conservation of that process is tested on its independently supplied pairing, as in Eq.~\eqref{eq:source-transport}.

\paragraph{Fundamental nonunitary source laws.}
An independently specified fundamentally nonunitary source can use this CP framework. A dilation does not establish a physically present conservative environment. For a complete output family, its unital dual map gives effects $E_r=\mathcal E(Q_r)$ with $\sum_rE_r=I$. Their probabilities still require the statistical conditions below.

A nonlinear normalized prescription instead requires its own preparation domain, treatment of zero-success inputs, retained labels and composite law. A conditional update after selection is not thereby a deterministic channel.

A constructive finite certificate is available directly on source Gram forms.  Suppose the source supplies positive semidefinite forms \(G_i,G_f\) and branch maps \(A_r\) with
\begin{equation}
\sum_r A_r^\dagger G_fA_r=G_i.
\label{eq:II-nonunitary-source-certificate}
\end{equation}
For \(x\in\ker G_i\), the sum of the nonnegative quantities \(\langle A_rx,G_fA_rx\rangle\) vanishes, so every \(A_rx\) lies in \(\ker G_f\).  The maps therefore descend to the positive quotients.  In orthonormal quotient coordinates their maps \(K_r\) satisfy \(\sum_rK_r^\dagger K_r=I\), giving a trace-preserving CP map and its ordinary composition law.  An inverse process is not required for this finite certificate. This certificate assumes the positivity of the supplied forms. Its quotient normalization is equivalently $W^\dagger W=I$ for the mathematical stacked map $W\psi=\sum_r|r\rangle\otimes K_r\psi$. A Kraus label still requires physical registration to become a readable record.  If a further Gauss sector is required, each branch must preserve it. Covariance of the summed channel alone is insufficient.

\paragraph{Mathematical adjoint versus physically constructible reversal.}
If the source contains an admitted inverse process \(N:\Sigma_f\to\Sigma_i\) with \(N\circ M=\operatorname{id}\) and \(M\circ N=\operatorname{id}\), functorial representation gives
\begin{equation}
T_NT_M=I,
\qquad
T_MT_N=I,
\qquad
T_N=T_M^{-1}.
\label{eq:II-physical-inverse-implies-operator-inverse}
\end{equation}
The converse does not follow.  The Hilbert adjoint or inverse of \(T_M\) can exist even when no admitted source process or available apparatus implements it.  The dagger in the primitive category encodes orientation reversal of a source process and the adjoint bridge. Constructible reversal is an additional physical property of the admitted controller class.

\section{How an interaction produces a record}
\label{sec:II-records-derived-complete}
When does a correlation become a record? A record cannot be defined solely by what the future apparatus fails to reveal.  Consider a qubit which reaches a cut \(\Sigma\) in
\begin{equation}
|+\rangle=\frac{|0\rangle+|1\rangle}{\sqrt2}
\label{eq:II-false-record-state}
\end{equation}
without any detector or marker interaction.  Suppose the declared future experiment performs only a readout of the Pauli observable \(Z:=|0\rangle\langle0|-|1\rangle\langle1|\).  At \(\Sigma\), the projections \(Q_0=|0\rangle\langle0|\) and \(Q_1=|1\rangle\langle1|\) commute with every future-readable observable of that fixed experiment and pull back from the later \(Z\) record.  Retention/inaccessible-coherence conditions alone would therefore label \(Q_0,Q_1\) as an earlier record although nothing physical happened at \(\Sigma\).  This false positive shows that record formation requires a past-directed production condition before a future-directed retention condition.

\begin{figure}[!htbp]
\centering
\begin{tikzpicture}[>=Latex,font=\small,
  state/.style={draw,rounded corners=1.5pt,align=center,text width=7.55cm,minimum height=.85cm,inner sep=3pt}]
\node[state] (ready) {Ready detector\\$(a|0\rangle+b|1\rangle)|D_*\rangle$};
\node[state,below=6mm of ready] (write) {Correlated detector labels\\$a|0,D_0\rangle+b|1,D_1\rangle$};
\node[state,below=6mm of write] (copy) {A readable copy in a fresh memory\\$a|0,D_0,D'_0\rangle+b|1,D_1,D'_1\rangle$};
\draw[->] (ready) -- node[right,font=\scriptsize] {writing interaction} (write);
\draw[->] (write) -- node[right,font=\scriptsize] {readout interaction} (copy);
\end{tikzpicture}
\caption{A reference example of writing and reading an orthogonal detector label, with $|a|^2+|b|^2=1$. The ready state is distinct from either written label. The complete state still has both components. Retention requires that the labels remain readable under the admitted later operations. A coherent erasure can remove the stored distinction and lies outside that retention requirement. These interactions construct record alternatives, without selecting one of them.}
\label{fig:writing-reading}
\end{figure}
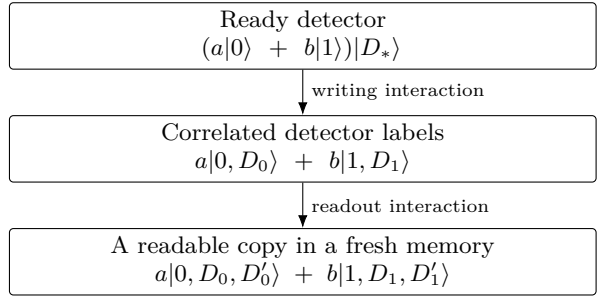
\subsubsection*{Physical registration of a record}
Before \(\Sigma\) can support a record family, the actual source history must contain a registration region from an earlier boundary \(\Sigma_-\),
\begin{equation}
R:\Sigma_-\longrightarrow\Sigma
\label{eq:II-registration-region}
\end{equation}
that creates a physically typed record algebra or readout interface.  In a factorizing representation the readability data consist of a record carrier \(D\), ready support \(P_*^D\), a source-derived registration map \(W_{\rm reg}\), and nonzero orthogonal record sectors \(P_r^D\).  Let \(P_-\) denote the reachable input support of the registration region and \(I\) the identity on the nonmemory factor.  Actual production remains a separate source condition.  The following relations specify orthogonal readable output support
\begin{equation}
\begin{gathered}
P_r^DP_s^D=\delta_{rs}P_r^D,\\
\left(\sum_r I\otimes P_r^D\right)W_{\rm reg}P_-=W_{\rm reg}P_-.
\end{gathered}
\label{eq:II-record-production}
\end{equation}
The sectors must be physically readable.  A transparent sufficient witness is an admitted readout/copy interaction into a fresh carrier \(D'\),
\begin{equation}
W_{\rm read}P_r^D
=(P_r^D\otimes P_r^{D'})W_{\rm read}
\label{eq:II-record-copy-witness}
\end{equation}
on the reachable record support, with \(P_r^{D'}P_s^{D'}=\delta_{rs}P_r^{D'}\). Here $W_{\rm read}$ includes insertion of the fixed ready state of $D'$ and acts from $D$ into $D\otimes D'$. Identity factors on any other degrees of freedom are suppressed. For this sufficient witness require \(W_{\rm read}^{\dagger}W_{\rm read}=P_{\rm reach}\) on that support, with every claimed record label reachable. This normalization excludes a zero or label-erasing readout map.  A nondestructive literal copy is not required by the definition. Any source-derived interface whose terminal readout sectors reliably distinguish the labels is sufficient.  In a nonfactorizing algebraic sector, Eq.~\eqref{eq:II-record-copy-witness} is replaced by an injective source-derived readout of a nontrivial commutative record algebra into terminal distinguishable sectors.

\paragraph{A sufficient ready-to-written witness.}
One explicit realization makes the production condition independently checkable. Figure~\ref{fig:writing-reading} shows the simple nondemolition case. Let the record carrier have orthogonal ready and written supports and require
\begin{equation}
\begin{gathered}
P_{\rm ready}^DP_{\rm written}^D=0,\qquad P_r^D\le P_{\rm written}^D,\\
(I\otimes P_{\rm ready}^D)P_-=P_-,\\
(I\otimes P_{\rm written}^D)W_{\rm reg}P_-=W_{\rm reg}P_- .
\end{gathered}
\label{eq:II-ready-written-witness}
\end{equation}
These flag conditions exclude the identity map, but an unrelated flag flip alone does not certify writing a specified label. For an explicit sharp-input witness prepare a fresh ready state, write \(T=W_{\rm reg}\iota\), \(\iota\psi=\psi\otimes|D_*\rangle\), require \(T^\dagger T=I\), and fix an input projection family \(\{A_r\}\) before choosing the interaction. Require, on the whole admitted input space,
\begin{equation}
(I\otimes P_r^D)T=TA_r,\qquad \sum_rA_r=I,
\label{eq:II-input-memory-intertwiner}
\end{equation}
with at least two nonzero branches when a nontrivial alternative is claimed. This correlates the memory with the specified input label, and allows destructive transfer of that label into the memory. A coherent swap can satisfy this relation. It transfers the information faithfully but does not establish decoherence, permanence or one actual outcome. If the writer is additionally claimed to preserve the system alternative, require both intertwiners:
\begin{equation}
(A_r\otimes I)T=TA_r=(I\otimes P_r^D)T.
\label{eq:II-nondemolition-writing}
\end{equation}
Equality between the two output expressions alone does not establish their relation to the input alternative. For a complete CP writing map, impose the applicable identities on each ready-input Kraus map \(T_\alpha\), with \(\sum_\alpha T_\alpha^\dagger T_\alpha=I\). Then the memory identity gives the effect \(\sum_\alpha T_\alpha^\dagger(I\otimes P_r^D)T_\alpha=A_r\). For a general noisy instrument, the source and readout can determine nonprojective effects. Such an instrument lies outside this stronger sharp-input witness.

A written inconclusive or no-click label is distinct from readiness. A deadline record therefore requires a writing or latching stage. This witness is sufficient rather than necessary. Its couplings, readout and retention must be physically supplied.

\paragraph{Quantitative writing and readback errors.}
Let \(T\) be an isometric writer on the ready input space, \(A_r\) a
specified input alternative, and \(Q_r=I\otimes P_r^D\). Suppose
\begin{equation}
\|Q_rT-TA_r\|\le\varepsilon_r,\qquad
\|\Lambda^*(Q_r^f)-Q_r\|\le\eta_r,
\label{eq:II-writing-readback-defects}
\end{equation}
where \(\Lambda^*\) is the unital CP pullback of a specified continuation.
For \(F_r=T^\dagger\Lambda^*(Q_r^f)T\),
\begin{equation}
\|F_r-A_r\|\le\varepsilon_r+\eta_r.
\label{eq:II-writing-retention-error}
\end{equation}
The proof is the identity
\[
F_r-A_r=
T^\dagger[\Lambda^*(Q_r^f)-Q_r]T+
T^\dagger(Q_rT-TA_r).
\]
Moreover, \(A_r\psi=\psi\), \(\|\psi\|=1\), implies
\(\|(I-Q_r)T\psi\|^2\le\varepsilon_r^2\).
A complete CP writer obeys the same bounds after its Kraus maps are stacked into \(T\), with the readout tensored with the Kraus identity. This is a mathematical representation, not an extra physical memory. If an actual initial state differs in trace distance by at most \(d_{\rm ready}\) from the stipulated ready preparation, each final effect expectation changes by at most \(d_{\rm ready}\), provided the same complete normalized map is defined on both initial states. This follows from trace-distance contraction and the effect bound. These operator estimates become probability-error estimates only after the statistical interpretation of the probability analysis. They quantify errors. They do not establish that an unconstructed apparatus meets the specified tolerances.

\paragraph{Scope of exactness.}
Exact flags and exact retention are properties of a specified model and admitted continuation class, not universal conclusions about relativistic detectors. A lower-bounded Hamiltonian alone does not exclude exact persistence: a projection commuting with that Hamiltonian is conserved exactly. A stricter local-field requirement can fail. If the vacuum is separating for a local algebra, as under the Reeh--Schlieder hypotheses, a local projection \(P\) with \(P\Omega=0\) must vanish~\cite{FewsterRejzner2019}. A nonzero strictly local click projection therefore cannot be perfectly silent in that vacuum. This does not identify a prepared ancillary memory with the vacuum, or rule out every exact finite-register model. The centrality result below also applies beyond type-I algebras. A factor has scalar center, while its restricted readable algebra can have a nontrivial center.

\paragraph{A mixed microscopic ready state is sufficient.}
Let $\{P_r\}$ be an orthogonal resolution of the reachable input support, $K_*$ a ready memory subspace, and $K_r$ mutually orthogonal written subspaces, also orthogonal to $K_*$ in the flagged construction. Suppose the source supplies isometries $W_r:K_*\to K_r$ and, on the ready support, the writing map
\[
 V=\sum_rP_r\otimes W_r.
\]
Then $V^\dagger V=I\otimes I_{K_*}$. For any positive trace-one memory state $\tau$ supported in $K_*$, not necessarily pure,
\[
 V(\rho\otimes\tau)V^\dagger
 =\sum_{r,s}P_r\rho P_s\otimes W_r\tau W_s^\dagger.
\]
With $Q_r=I\otimes\Pi_{K_r}$, projection and partial trace give
\[
\begin{aligned}
 \mathcal I_r(\rho)&=\operatorname{Tr}_D[Q_rV(\rho\otimes\tau)V^\dagger Q_r]\\
 &=P_r\rho P_r,\qquad E_r=P_r.
\end{aligned}
\]
Thus the selective maps and effects are independent of the microscopic blend within the ready subspace. A pure apparatus state is unnecessary for this coupling model. The stated registration and retention conditions still apply.

\subsubsection*{Cut invariance through record-free reversible regions}
Let a source isomorphism have a unitary represented map \(U:\mathcal H_{\Sigma_0}\to\mathcal H_{\Sigma_1}\), and pass through a region containing no registration interaction for the record interface under consideration.  Identify the same record, readout and admitted continuation data at the two cuts by this transport, and let \(\mathfrak D_{\Sigma_j}\) denote the corresponding record algebra at boundary \(\Sigma_j\).  Then the record algebra and its projections are merely transported,
\begin{equation}
\mathfrak D_{\Sigma_1}=U\mathfrak D_{\Sigma_0}U^\dagger,
\qquad
Q_r^{(1)}=UQ_r^{(0)}U^\dagger.
\label{eq:II-record-cut-invariance}
\end{equation}
Unitary conjugation is a \(*\)-isomorphism and preserves orthogonality, completeness and every transported readout relation.  Mere invertibility of a represented source map would not suffice: \(U=\operatorname{diag}(1,1/2)\) sends the projector \(Q=|1\rangle\langle1|\) to \(UQU^\dagger=Q/4\), which is not a projector.  More generally one may supply an independent normal \(*\)-isomorphism of the record algebras, together with the same operational transport, instead of this unitary implementation.  Consequently
\begin{equation}
\boxed{
\mathfrak D_{\Sigma_0}\ \text{is nontrivial}
\Longleftrightarrow
\mathfrak D_{\Sigma_1}\ \text{is nontrivial}.}
\label{eq:II-cut-invariance-theorem}
\end{equation}
In particular a ready/trivial record algebra cannot become a record algebra simply because the mathematical cut is moved.  The beginning of a record is tied to an actual registration subregion.

\subsubsection*{Retention of an already produced record}
Once the record has been written, what operations can the experiment still perform on it? Fix a registered record family at \(\Sigma\).  Let \(\mathrm{Cont}_{\mathfrak E}(\Sigma)\) be the continuation class declared by the physical source arrangement.  For each \(\Lambda\) in that class, let \(\mathfrak N_{f_\Lambda}\) and \(\mathfrak N_\Sigma\) denote, respectively, the terminal and intermediate readable algebras, and let
\begin{equation}
\Lambda^*_{\Sigma\leftarrow f_\Lambda}:
\mathfrak N_{f_\Lambda}
\longrightarrow
\mathfrak N_\Sigma
\end{equation}
be the Heisenberg pullback, and let \(P_{\rm rec}^\Sigma\) be the reachable support on which the record family is complete.  The produced records are retained records when
\begin{align}
Q_r^\Sigma Q_s^\Sigma&=\delta_{rs}Q_r^\Sigma,
&\sum_rQ_r^\Sigma&=P_{\rm rec}^\Sigma,\\
\Lambda^*_{\Sigma\leftarrow f_\Lambda}(Q_{r,\Lambda}^{f_\Lambda})&=Q_r^\Sigma,
\label{eq:II-record-retention}\\
Q_r^\Sigma A Q_s^\Sigma&=0\qquad(r\ne s)
\label{eq:II-record-offdiagonal}
\end{align}
for every admitted continuation and every later-readable \(A\in\mathfrak N_\Sigma\), with \(P_{\rm rec}^\Sigma\) reducing the readable algebra. The continuation class is specified before the outcome and includes every alternative used in a comparison. It is not chosen afterward to retain only favourable histories.  Appending a new recohering controller not included in the declared continuation class defines a different enlarged source process.

For a complete finite record family on that support, let \(\mathcal D_Q(X)=\sum_rQ_rXQ_r\).  Trace duality and separation by density operators give the exact finite-dimensional criterion
\begin{equation}
\begin{gathered}
\operatorname{Tr}[(\rho-\mathcal D_Q\rho)E]=0\quad\hbox{for all }\rho\\
\Longleftrightarrow\quad E=\mathcal D_Q(E).
\end{gathered}
\label{eq:II-record-operational-domain}
\end{equation}
For Hermitian \(E\), the maximum absolute discrepancy over all states is \(\|E-\mathcal D_Q(E)\|\).  Restricting the reachable states can weaken the necessity direction.  Thus invariance on one declared readable algebra does not establish equivalence after arbitrary new coherent controls are admitted.

On the reducing support the off-diagonal condition gives
\begin{equation}
[Q_r^\Sigma,A]=0,
\end{equation}
The readback condition also puts each \(Q_r^\Sigma\) in the intermediate readable algebra.  Combining that membership with commutation gives
\begin{equation}
\boxed{
\operatorname{vN}\{Q_r^\Sigma\}
\subseteq
\operatorname{Cent}\!\left(P_{\rm rec}^\Sigma
\mathfrak N_\Sigma
P_{\rm rec}^\Sigma\right).}
\label{eq:II-record-center-inclusion}
\end{equation}
Centrality is therefore a consequence of produced and retained storage. It cannot replace the registration criterion.  Approximate decoherence or approximate centrality can support quantitative retention after a record has been produced, but neither creates registration nor selects one realized outcome.

For a terminal retained-record family \(\{Q_r^f\}\) complete on the range of a complete-process isometry \(V_{\mathfrak E}:\mathcal H_{\rm in}\to\mathcal H_f\), let \(P_{\rm rec}^f=\sum_rQ_r^f\) and require \(P_{\rm rec}^fV_{\mathfrak E}=V_{\mathfrak E}\).  The preparation-side effects---the operators Englert calls probability operators once the statistical interpretation is supplied---are then
\begin{equation}
\boxed{
E_r:=V_{\mathfrak E}^\dagger Q_r^fV_{\mathfrak E},
\qquad E_r\ge0,
\qquad \sum_rE_r=I.}
\label{eq:II-universal-record-effects}
\end{equation}
No probability interpretation is used in this pullback.

Writing a record, keeping it readable and observing one result are thus separate physical questions. The interference examples in Sec.~\ref{sec:I-paradoxes} show why loss of visibility alone answers none of them.
\label{eq:II-marker-registration-retention-occurrence}

\label{sec:II-universal-theorem}
Figure~\ref{fig:source-record-probability} summarizes the construction. The source conditions determine the representation, the apparatus supplies readable alternatives, and retention specifies their persistence. The remaining question is their probability.

\section{Records and probabilities}
\label{joint:probabilities}
\subsection{Source and apparatus inputs to the probability theorem}
\label{sec:I-upstream}
\label{sec:I-types-complete}
Write $V$ for the complete isometric apparatus of Eq.~\eqref{eq:II-universal-record-effects}. Its terminal projections $Q_r$ give
\begin{align}
 V^\dagger V&=I,\label{eq:I-V-source-derived}\\
 \sum_r Q_r&=P_{\rm rec},\qquad P_{\rm rec}V=V,
 \label{eq:I-record-support-complete}\\
 B_r&=Q_rV,\qquad E_r=B_r^\dagger B_r,
 \label{eq:I-Br-Er-complete}\\
 E_r&\ge0,\qquad\sum_rE_r=I.
 \label{eq:I-effect-normalization}
\end{align}
The first identity preserves every length and overlap. Adjoint reversal alone does not imply it: a half-power attenuator has $V=I/\sqrt2=V^\dagger$ but $V^\dagger V=I/2$. Conservation of the physical pairing, or a represented reversal that undoes the process, is the additional condition. A complete nonunitary source uses $E_r=\mathcal E(Q_r)$ instead. Neither construction yet assigns chances to its alternatives.

\subsection{Complete later refinements preserve the earlier record effect}
\label{sec:I-later-refinement}
Fix an already stored record \(r\).  Let
\(C_{s\alpha}^{(r)}:Q_r\mathcal H_{\rm out}\to\mathcal H_{\rm later}\)
be Kraus maps of a complete later source-derived instrument acting inside that record sector, where \(s\) labels the later stored record and \(\alpha\) is the Kraus index.  Completeness means
\begin{equation}
\sum_{s,\alpha}
C_{s\alpha}^{(r)\dagger}C_{s\alpha}^{(r)}
=Q_r.
\label{eq:I-future-completeness}
\end{equation}
Define
\begin{align}
B_{r,s,\alpha}&:=C_{s\alpha}^{(r)}B_r,\\
G_{r,s}&:=\sum_\alpha
B_{r,s,\alpha}^\dagger B_{r,s,\alpha}.
\label{eq:I-refined-effect}
\end{align}
Then
\begin{align}
\sum_sG_{r,s}
&=B_r^\dagger
\left(\sum_{s,\alpha}C_{s\alpha}^{(r)\dagger}C_{s\alpha}^{(r)}\right)B_r\\
&=B_r^\dagger Q_rB_r
=B_r^\dagger B_r
=E_r.
\label{eq:I-later-refinement-identity}
\end{align}
Summing over all later outcomes gives the operator for the earlier stored record. This equality uses completeness of the later instrument and retention of that record. It holds before probabilities are assigned.

\subsection{What statistical property must the preparation have?}
\label{sec:preparation-sufficiency}
The detector's record operator is a mathematical consequence of its represented coupling. Its probability is a further physical question. Suppose two preparation procedures, $\alpha$ and $\beta$, give the same represented state. Must the same subsequent reader give the same probabilities? An omitted memory can make the answer no.

To state the issue without using Born's rule, let $x$ denote the complete initial physical data. A source with a specified statistical law supplies preparation measures $\nu_\alpha$ and reader response kernels $k_D(B|x)$. Here $D$ names the complete experiment and $B$ an event in its record. Its actual law is
\begin{equation}
 p_D(B|\alpha)=\int k_D(B|x)\,\nu_\alpha(dx).
 \label{eq:actual-reader-law}
\end{equation}
For deterministic dynamics, $k_D(B|x)$ is the indicator that the resulting record lies in $B$. Random initial data still require a specified measure. A stochastic source instead supplies its noise and response law.

The represented state $\rho_\alpha$ is reconstructed independently. Preparation sufficiency for this reader means
\begin{equation}
\begin{split}
 \rho_\alpha=\rho_\beta\quad\Longrightarrow\quad
 \int k_D(B|x)\,[\nu_\alpha-\nu_\beta](dx)=0.
\end{split}
\label{eq:preparation-sufficiency}
\end{equation}
This condition does not define $\rho$ by equality of measured probabilities. It asks whether a state supplied by the source omits anything that can change them. One sufficient dynamical condition is that the reader depends only on a specified function $y(x)$ and that equal represented preparations induce the same law of $y$. Both statements then require a proof about that source.

Appendix~\ref{new:history-appendix} gives an exact finite test of this condition, a measure of its failure and a delayed-memory counterexample. Those results test an independently specified physical history law. They are not prerequisites for the short uniqueness argument that follows, and a finite-state model does not supply its full preparation domain.

\subsection{Born uniqueness from preparation and reader calibration}
\label{sec:I-born-main}
\subsubsection{The physical and statistical inputs}
\label{sec:I-main-born-assumptions}
Fix a terminal yes/no measurement readout (abbreviated below as the reader) with sharp projection $Q$ on its complete represented input. Three questions about that reader will suffice. Does its represented state contain all preparation information relevant to the reading? Can preparations be mixed by an independent classical choice? Does the reader distinguish its yes and no sectors with certainty?
\begin{description}
\item[Preparation sufficiency] Each admitted procedure with the same represented density operator (statistical operator, in Englert's terminology) gives the same yes probability $q(\rho)$. This is Eq.~\eqref{eq:preparation-sufficiency} for the selected reader.
\item[Independent fair mixing] A classical selector can choose either of any two admitted procedures with probability $1/2$, independently of their subsequent reader responses. Ignoring the selector gives the represented midpoint $(\rho+\sigma)/2$. The selected conditional responses are those of the original procedures. Ordinary total probability therefore gives
\begin{equation}
 q\!\left(\frac{\rho+\sigma}{2}\right)
 =\frac{q(\rho)+q(\sigma)}2.
 \label{eq:reader-midpoint}
\end{equation}
The fairness and independence of this selector are statistical inputs. They are not certified by assigning Born probabilities to an auxiliary superposition. A fixed independently calibrated bias strictly between zero and one also suffices, as Appendix~\ref{app:representation-details} shows.
\item[Sharp reader calibration] A preparation wholly in the range of $Q$ gives yes with certainty. A preparation in its kernel gives no with certainty. The stated preparation family contains the real two-dimensional disk used below, including its mixtures. Actual probabilities lie between zero and one.
\end{description}
These are conditions on complete preparation and readout procedures, including any correlated side records. They are weaker in domain than demanding a probability function on every effect in every Hilbert space, but they remain substantive statistical and physical premises. Finite experimental agreement with them is evidence over the settings tested. An exact theorem over the disk requires them throughout that disk.

\subsubsection{The uniqueness argument}
\label{sec:I-plain-born-proof}
Let $\psi$ be a normalized represented preparation and put $a=\langle\psi,Q\psi\rangle$. Calibration settles $a=0$ and $a=1$. For $0<a<1$, define orthonormal vectors
\[
\begin{gathered}
 u=\frac{Q\psi}{\sqrt a},\qquad v=\frac{(I-Q)\psi}{\sqrt{1-a}},\\
 \psi=\sqrt a\,u+\sqrt{1-a}\,v.
\end{gathered}
\]
On their two-dimensional span let $Z=|u\rangle\langle u|-|v\rangle\langle v|$ and $X=|u\rangle\langle v|+|v\rangle\langle u|$. The required real disk is
\begin{equation}
 \rho(x,z)=\tfrac12(I+xX+zZ),\qquad x^2+z^2\le1.
 \label{eq:reader-disk}
\end{equation}
Here $I$ is the identity on this span. The domain condition requires actual preparations represented by every point of the disk, with midpoint mixing inside it. No assertion that every formal Hilbert vector is physically preparable is used.

\begin{figure}[!t]
\centering
\begin{tikzpicture}[>=Latex,font=\small,x=1.30cm,y=1.30cm]
\fill[black!3] (0,0) circle (1);
\draw (0,0) circle (1);
\draw[->] (-1.18,0)--(1.2,0) node[right] {$x$};
\draw[->] (0,-1.17)--(0,1.2) node[above] {$z$};
\draw[thick] (-.8,.35)--(.7,.35);
\fill (-.8,.35) circle (1.4pt);
\fill (.7,.35) circle (1.4pt);
\fill (-.05,.35) circle (1.6pt);
\node[below,font=\scriptsize] at (-.05,.35) {midpoint};
\fill (0,1) circle (1.5pt);
\fill (0,-1) circle (1.5pt);
\node[anchor=west,font=\scriptsize] at (.12,1.02) {$u$: certain yes};
\node[anchor=west,font=\scriptsize] at (.12,-1.02) {$v$: certain no};
\node[align=left,anchor=west,font=\small] at (1.52,.35)
 {$f(x,z)=\frac12+b_xx+\frac12z$\\[5pt]
  $\min f=\frac12-\sqrt{b_x^2+\frac14}$\\[5pt]
  $f\ge0\ \Longrightarrow\ b_x=0$};
\end{tikzpicture}
\caption{The preparation disk used for one terminal reader. Physical mixing gives the midpoint relation on every chord. Bounded probabilities extend it to an affine function on the disk. Calibration fixes its values at the two poles. Any nonzero transverse coefficient would make the probability negative somewhere on the disk. The resulting law is $f(x,z)=(1+z)/2$. Availability of the disk preparations is an explicit physical condition.}
\label{fig:born-disk}
\end{figure}
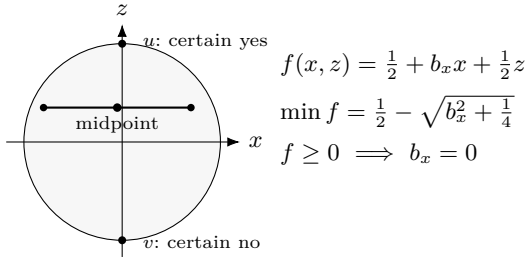

Write $f(x,z)=q(\rho(x,z))$. Equation~\eqref{eq:reader-midpoint} makes $f$ midpoint affine. Its bound $0\le f\le1$ makes it fully affine without an extra continuity assumption. To see this on any segment, subtract its endpoint interpolation and call the result $e(t)$. For $0\le t\le1/2$ midpoint affinity gives $e(t)=e(2t)/2$. For $1/2\le t\le1$ it gives $e(t)=e(2t-1)/2$. Iteration bounds $|e(t)|$ by a fixed constant times $2^{-n}$, so $e(t)=0$. Segment affinity on a disk with nonempty interior is ordinary affine linearity. Thus
\[
 f(x,z)=c+b_xx+b_zz.
\]
Calibration gives $f(0,1)=1$ and $f(0,-1)=0$, hence $c=b_z=1/2$. Positivity on the whole disk forces
\[
 \sqrt{b_x^2+1/4}\le1/2,
\]
so $b_x=0$. Therefore
\begin{equation}
 f(x,z)=\frac{1+z}{2},\qquad q(|\psi\rangle\langle\psi|)=a.
 \label{eq:reader-born}
\end{equation}
Figure~\ref{fig:born-disk} shows the geometry of this step. The pure state $\psi$ has disk coordinates $x=2\sqrt{a(1-a)}$ and $z=2a-1$. The probability of its yes record is consequently $\langle\psi,Q\psi\rangle$.

An independently represented complete apparatus sends an input state to the terminal reader. For an isometry $V$ and sharp readout $Q_r$, the theorem gives
\begin{equation}
\begin{gathered}
 p(r|\psi)=\|Q_rV\psi\|^2=\langle\psi,E_r\psi\rangle,\\
 E_r=V^\dagger Q_rV.
\end{gathered}
 \label{eq:I-born-pure-main}
\end{equation}
This application requires the disk conditions at that terminal reader. The range of $V$ can be smaller than the required preparation disk. The extra preparations and controls therefore have to be physically supplied, rather than inferred from the existence of $V$. A mixed terminal state, or a fundamentally nonunitary complete process, uses the full preparation extension proved in Appendix~\ref{app:full-reader}, or an actual admitted mixture for which the pure-state result applies. A mathematical dilation alone does not supply those physical preparations.

For a normal trace-preserving represented process $\Phi$ whose output preparations satisfy that extension,
\begin{equation}
 p(r|\rho)=\operatorname{Tr}[\Phi(\rho)Q_r]
          =\operatorname{Tr}[\rho\Phi^*(Q_r)].
 \label{eq:born-general-process}
\end{equation}
The effects of different experiments can be noncommuting. Even one generalized measurement can have noncommuting effects. Probability addition concerns mutually exclusive detector records in the selected experiment, not simultaneous sharp values of incompatible observables.

\paragraph{Relation to complete-event comparisons.}
The proof does not assume a universal function of branch length. It derives the trace formula from preparation sufficiency, independent mixing and reader calibration. Complete experiments yielding the same effect expectation then have the same probability, provided each satisfies those conditions. The alternative comparison-and-splitting characterization is given separately in Appendix~\ref{app:comparison}. It is not an additional premise of this proof.

The mathematical uniqueness statement belongs to the family of positive-functional characterizations of quantum probabilities, including Gleason's, Busch's and the effect-frame formulation of Caves and collaborators~\cite{Gleason1957,Busch2003,CavesFuchsManneRenes2004}. The distinction here is the explicit connection to source preparations, physical registration, retained histories and an independently testable preparation-sufficiency condition. This is not an assumption-free replacement for those theorems. The local proof identifies a sufficient preparation domain for each pure target. An exact source derivation of its statistical inputs remains a separate physical result.

\paragraph{Why preparation sufficiency carries statistical content.}
Consider the state $\operatorname{diag}(1/3,2/3)$ in the $u,v$ basis. It can be prepared by a classical mixture of $u,v$ with weights $1/3,2/3$, or an equal mixture of
\[
 \psi_\pm=\sqrt{1/3}\,u\pm\sqrt{2/3}\,v.
\]
A hypothetical normalized-square rule assigns yes probability $1/5$ to each $\psi_\pm$, while sharp calibration and total probability give $1/3$ for the first preparation. The same represented density operator would have two reader probabilities. Preparation sufficiency excludes that assignment. It is a condition with physical content, not an algebraic identity.

Likewise sharp calibration alone is insufficient. On the disk,
\begin{equation}
 f_\epsilon(x,z)=\frac{1+z}{2}+\epsilon x^2,
 \qquad 0<\epsilon\le\frac14,
 \label{eq:reader-nonaffine}
\end{equation}
is a probability between zero and one with the correct endpoint calibrations. It fails midpoint mixing. The disk inequality $x^2\le1-z^2$ proves its upper bound, and its other terms prove the lower bound. Thus the exact place where statistics enters is visible and can be tested.

\subsection{Can causality explain preparation sufficiency?}
\label{sec:causality-route}
An alternative route starts with record laws for pure composite preparations~\cite{Nenashev2016}. Suppose the law depends only on the represented ray for a fixed apparatus. Assume no-signalling and invariance when an idle auxiliary system is added or transported. Pure preparations with the same reduced state are related by a unitary on a sufficiently large distant partner. When that operation is physically admitted, these assumptions give the same local reader law. This is preparation sufficiency for that class, not a consequence of the partial trace alone.

To fix the value, assume ordinary positive normalized probabilities for one retained joint record per run, and that a joint record of zero represented amplitude never occurs. With the auxiliary preparations, readable copies and controls of Appendix~\ref{app:causal-born}, equal branches have equal chances. Rational comparisons and joint-record inclusions then give
\begin{equation}
 \frac{\lfloor Ma\rfloor}{M}\le p(r)
 \le\frac{\lceil Ma\rceil}{M},
 \label{eq:causal-brackets}
\end{equation}
where $a$ is the squared branch weight. Comparisons for every $M$ force $p(r)=a$ without continuity. For a fixed $M$, $|p(r)-a|\le1/M$ bounds the comparison with exact controls, not the energy, duration or experimental error.

The appendix treats arbitrary-rank record sectors and specifies retention under every continuation used in a comparison. Classical mixtures additionally require independent selectors with the stated conditional responses. This route replaces the disk assumptions with completeness, locality, calibration and control assumptions. It does not derive their physical origin or infer probabilities from causality alone.

\subsection{Normalization and physical mixed preparations}
The following consequences apply whenever either probability route
holds on the relevant preparation and reader domain. We now separate
normalization of a representative from conservation, and then consider
mixtures actually produced by the apparatus.
\paragraph{Unit norm and conservation are different statements.}
For a nonzero representative of the same pure preparation, the result is
\begin{equation}
 p(E|\psi)=\frac{\langle\psi,E\psi\rangle}{\langle\psi,\psi\rangle}.
 \label{eq:I-homogeneous-born}
\end{equation}
Choosing a unit representative is a convention. The dynamical assertion $V^\dagger V=I$ preserves every norm and overlap in an independently specified pairing. Its source conditions were stated before probability. Statistical normalization on a complete record set also differs from a guarantee that a detector will click by a chosen deadline. A written no-click result can be part of that complete set.

\paragraph{Mixed preparations generated by physical sources.}
An actual preparation label $j$ with probability $w_j$ and conditional state $\rho_j$ gives, on ignoring that label,
\begin{equation}
 p(E)=\sum_jw_jp(E|\rho_j).
 \label{eq:I-physical-blend-step}
\end{equation}
When each preparation and readout is in the theorem's domain, the pure-state result extends to
\begin{equation}
 p(E|\rho)=\operatorname{Tr}(\rho E),\qquad
 \rho=\sum_jw_j\rho_j.
 \label{eq:I-born-mixed-main}
\end{equation}
The weights are previously derived record probabilities or independently specified classical preparation probabilities. An arbitrary mathematical decomposition does not supply them. Countable blends require a countably additive preparation law. Their trace-norm convergence follows from the tail identity $\|\sum_{j>m}w_j\rho_j\|_1=\sum_{j>m}w_j\to0$.

For an actual composite $SR$, the joint theorem likewise gives
\begin{equation}
\begin{split}
 p(E)&=\operatorname{Tr}_{SR}[\rho_{SR}(E\otimes I_R)]\\
     &=\operatorname{Tr}_S[(\operatorname{Tr}_R\rho_{SR})E].
\end{split}
\label{eq:I-actual-composite-restriction}
\end{equation}
A mathematical purification does not establish a physically present auxiliary system. These physical preparations and restrictions give the same predictions for the same density operator on their admitted domain. Retained preparation labels and correlations can distinguish enlarged preparations, as the next example shows.

\paragraph{The same single-photon state can hide different pair correlations.}
An equal preparation mixture of the two $Z$ eigenstates has the same single-qubit density matrix $I/2$ as an equal mixture of the two $X$ eigenstates. If each preparation supplies two identical copies, their pair states are instead
\[
\begin{gathered}
 R_Z=\frac{I\otimes I+Z\otimes Z}{4},\\
 R_X=\frac{I\otimes I+X\otimes X}{4}.
\end{gathered}
\]
These are different: the $Z\otimes Z$ expectations are one and zero. If the two preparation labels are chosen independently, both give $(I/2)\otimes(I/2)$ instead. The correlation is additional physical information, not a choice of decomposition of the same complete state. Paired-state learning and its photon implementation use just this distinction~\cite{AgarwalPairs2025,DobneyPairs2026}. The experiment groups separately detected signal photons using preparation timing. The herald photon is not the second member of the learning pair. Its inference uses calibrated quantum detection probabilities. It therefore tests predictions after the statistical interpretation has been supplied. It is not an independent derivation of that interpretation.

\section{Recorded histories and individual outcomes}
\label{new:history-main}
The single-reader theorem fixes a probability once the apparatus and
its readout are specified. To follow several readings, we also need
the change of state produced by each interaction. This additional
information is contained in an instrument, the family of state maps
labelled by its possible records.

\subsection{Why the recorded interaction matters}
\label{sec:effect-instrument}
The unread evolution alone does not identify the record. A qubit illustrates this directly. The instruments with selective Kraus maps
\[
 \{P_0,P_1\}\quad\hbox{and}\quad\{I/\sqrt2,Z/\sqrt2\}
\]
both have unread channel
\[
 \rho\longmapsto P_0\rho P_0+P_1\rho P_1
                  =\tfrac12(\rho+Z\rho Z).
\]
For input $|0\rangle$, their record probabilities are respectively $(1,0)$ and $(1/2,1/2)$ after the probability theorem. A material readout determines which instrument is present. Equally, an effect determines an outcome probability but not the outgoing conditional state. The interactions $P_r$ and $U_rP_r$ have the same effects and generally different outputs.

\subsection{The probability law of a complete recorded history}
\label{sec:I-histories}
\subsubsection*{Joint probabilities of all recorded outcomes}
Let \(\mathfrak E\) denote the complete finite source process under consideration, and let $\mathcal P$ be the finite partially ordered set of its actual recording interactions. Its fixed classical causal order is denoted $\prec$. The common active state contains every memory or field degree of freedom that can affect a later record. At site $v$, the source-derived outcome map is $\mathcal I_{v,r}$. The sum over all records, including a written no-click label where needed, is required to preserve trace. For a complete conservative apparatus this follows from its isometry and exhaustive readout. For an independently specified completely positive source it is a normalization condition that must be established or supplied. Complete positivity alone does not give it. The represented apparatus couplings determine these maps when their source identification has been established. Listing independence uses the separate commutation condition for causally disjoint operations.

For this fixed experiment choose an admissible order of its recording sites and call them $1,\ldots,n$. Write $h=(r_1,\ldots,r_n)$ for a complete record. Then
\begin{align}
 \mathcal I_h&=\mathcal I_{n,r_n}\circ\cdots\circ\mathcal I_{1,r_1},\\
 G_h&=\mathcal I_h^*(I),\\
 p_\rho(h)&=\operatorname{Tr}\mathcal I_h(\rho)
           =\operatorname{Tr}(\rho G_h).
 \label{eq:I-history-law}
\end{align}
The star denotes the dual map defined by the trace pairing, and $\rho$ is the complete initial density operator. {The state maps act on trace-class operators. Their Heisenberg duals are normal maps on bounded observables.} The last line applies the probability theorem once to the final record, provided the complete history and its grouped events belong to the domain of the preparation and reader theorem. It does not assume a collapse at every factor.

If the full selective maps of causally disjoint couplings commute, every admissible listing gives the same complete output map. Summing the last record first gives normalization and consistency on each subset containing all causal predecessors of its interactions. Appendix~\ref{app:history-proof} proves these statements by exchanging adjacent incomparable operations and using preservation of the identity by the dual maps. They concern the same recorded experiment: marginalizing a stored reading is not removing its interaction.

The finite joint probabilities determine one law for the finite recorded history. The extension to countably many recording steps requires a regularity condition on each space of possible readings. This condition includes finite or countable labels and ordinary real-valued readings.\footnote{Precisely, the possible-reading spaces are standard Borel spaces: measurable spaces isomorphic to Borel subsets of complete separable metric spaces. A Borel set belongs to the collection generated from open sets by complements and countable unions. This condition permits the measure extension and conditional-distribution theorems used here. An arbitrary measurable space can fail to satisfy them.} If all finite joint laws agree when unwanted recorded coordinates are summed out, the Kolmogorov extension theorem gives one probability measure for the full history, on the measurable sets generated by questions about finitely many coordinates. An analogous extension for consistent positive operator-valued measures is available under its stated hypotheses~\cite{Tumulka2008}. Alternatively, a continuum history measure can be constructed directly, without an intermediate time grid. Finite numbers of events in compact regions must follow from that construction. Finite detector resolution alone does not prove this property.

For a past record $h$ with $p_\rho(h)>0$, define
\[
 \rho_h=\frac{\mathcal I_h(\rho)}{\operatorname{Tr}\mathcal I_h(\rho)}.
\]
For any specified subsequent recorded sequence $k$, the joint law gives
\begin{equation}
\begin{split}
 p(k|h)&=\frac{\operatorname{Tr}[(\mathcal I_k\circ\mathcal I_h)(\rho)]}
                   {\operatorname{Tr}\mathcal I_h(\rho)}\\
       &=\operatorname{Tr}\mathcal I_k(\rho_h).
\end{split}
\label{eq:I-conditional-history}
\end{equation}
Thus the conditional state is a way of calculating the conditional probabilities already fixed by the joint law. This conditional updating is state reduction in Englert's terminology; no additional physical collapse dynamics enters this identity. It explains neither why $h$ occurs nor which record a run produces.

For real-valued readings, regular conditional laws exist under the stated measure conditions and are fixed only almost everywhere. The ratio above is not defined at a zero-probability record. The full active state retains predictive memory. Detection-time records fix waiting-time laws when included, while a discrete schedule alone specifies no intervening continuous trajectory.

\subsubsection*{Covariance and independence of ordering}
To compare different experiments, restore the experiment label and write their laws as $p_{\rho,\mathfrak E}$. Let $\Lambda$ transform the preparation, apparatus, boundaries, couplings and record coordinates together. If the source maps transform covariantly, trace invariance gives
\begin{equation}
 p_{\rho_\Lambda,\Lambda\mathfrak E}(\Lambda B)
 =p_{\rho,\mathfrak E}(B).
 \label{eq:I-history-covariance}
\end{equation}
Only causal precedence is fixed. No choice of spacelike surfaces as successive instants is required. A listing may be used to calculate conditional probabilities, but the commutation argument makes the law independent of that listing. Covariance alone does not prove commutation of the complete operations for spacelike outcomes. It must follow from the source couplings~\cite{FewsterVerch2020}.

Spacelike commutation does not impose product probabilities on entangled preparations. For $|\Phi^+\rangle=(|00\rangle+|11\rangle)/\sqrt2$, take
\(A_x=\cos\theta_x\,Z+\sin\theta_x\,X\) and
\(B_y=\cos\varphi_y\,Z+\sin\varphi_y\,X\).  The sharp-record law is
\begin{equation}
 p(a,b\mid x,y)=\tfrac14[1+ab\cos(\theta_x-\varphi_y)],
 \quad a,b=\pm1.
 \label{eq:I-history-bell}
\end{equation}
It has unbiased distant marginals and retains the quantum correlations. It is a law for the records of the settings actually implemented, not a common value assignment to incompatible unperformed experiments.

\subsection{A probability law and one actual record}
\label{sec:actuality}
Fix the preparation and complete experiment. Write $P$ for its history law on the possible records $\Omega$. A measurable function $F$ of the complete record has average
\begin{equation}
 \mathbb E_P[F]=\int_\Omega F(h)\,P(dh).
 \label{eq:I-record-average}
\end{equation}
A specified record $h_*$ instead gives the value $F(h_*)$. These are different objects. For an event $B$ with $0<P(B)<1$, its indicator has variance $P(B)[1-P(B)]$ under the law and zero in one specified record. The formula for the law does not identify which $h_*$ occurs.

For finitely many records, every normalized positive functional on their commutative algebra is a unique mixture of point evaluations,
\[
 \mathbb E_P[F]=\sum_{h\in\Omega}P(h)F(h).
\]
This follows by evaluating the functional on the indicator of each individual record. It is a decomposition of an average, not a mechanism selecting one term. The empirical occurrence statement adopted here is that a completed run has one realized retained record~\cite{Englert2013}.

Equation~\eqref{eq:I-conditional-history} gives predictions after a known record without an additional collapse law. A physical collapse theory proposes dynamics beyond that identity and is tested through its history law. The algebraic closure does not exclude such dynamics.

A deterministic function of the same fixed preparation and apparatus data gives one fixed record. It cannot reproduce a nondegenerate distribution over repeated runs unless additional run-dependent data and their distribution are supplied. This observation does not exclude deterministic theories with such data. Nor does specifying a stochastic process explain its physical noise without a source law for that noise.

An unobserved variable can have different joint laws with the same record marginal. For example, $P(dh)\delta_0(d\xi)$ and $P(dh)\delta_1(d\xi)$ agree on every function of $h$. If $\xi$ becomes readable, it belongs to an enlarged experiment whose joint law requires a further physical construction. Predictive agreement on the admitted records is therefore distinct from a unique ontology.

A normalized deadline instrument can include a retained no-click result. If $E_C$ is the sum of successful effects, success by that deadline is certain for every represented input exactly when $E_C=I$. This follows by testing the positive operator $I-E_C$ on all vectors. It does not identify readiness with a written no-click record.

\paragraph{Repeated frequencies.}
For $N$ independently reset runs with $K$ outcomes, let $\Pi_{\rm bad}$ select strings whose empirical frequencies differ from the single-run squared branch weights by more than $\delta$. The product-amplitude calculation gives
\begin{equation}
 \|\Pi_{\rm bad}\Psi_N\|^2\le2K e^{-2N\delta^2}.
 \label{eq:I-frequency-concentration-example}
\end{equation}
Appendix~\ref{app:history-proof} gives the calculation and its probability interpretation. Before the Born theorem this is a norm estimate. After that theorem, and with independent resets, it is a probability estimate. It does not independently show that small-norm histories occur rarely~\cite{CavesSchack2005}.

\section{Coherence, individual records and predictive completeness}
\label{joint:completeness}

\subsection{Coherent processes, markers, records and the familiar paradoxes}
\label{sec:I-paradoxes}

We can now return to the interference experiment with which we began. What changes when a path is marked, when a detector writes a result, and when that result is used in a later prediction?

For a two-path interferometer, let \(V_{\rm fr}\) denote fringe visibility and \(D\) the optimized distinguishability stored in an admitted which-way marker.  The general duality relation obeys
\begin{equation}
D^2+V_{\rm fr}^2\le1,
\label{eq:I-englert-duality}
\end{equation}
with equality for the standard balanced pure two-path model~\cite{Englert1996}.  Neither quantity is a record criterion.  Indeed, the state
\begin{equation}
\rho_{QD}
=\frac12(|0\rangle\langle0|+|1\rangle\langle1|)\otimes\sigma_D
\label{eq:I-zero-visibility-zero-distinguishability}
\end{equation}
has \(V_{\rm fr}=0\) while the conditional detector states are identical and hence \(D=0\).  Absence of fringes therefore proves neither which-way information nor registration.  Conversely, a coherently reversible marker can have \(D>0\) without yet being a retained record.  Experiments on nonerasing quantum erasure provide the complementary warning: conditional high-visibility fringes can occur even where no which-way information was available to erase~\cite{SchwindtKwiatEnglert1999}.

Consider first two coherent alternatives, represented by vectors $\psi_1$ and $\psi_2$. Let $B_x$ map an incoming state to the outgoing amplitude associated with detector element $x$. Their squared amplitude satisfies
\begin{align}
\|B_x(\psi_1+\psi_2)\|^2
&=\|B_x\psi_1\|^2+\|B_x\psi_2\|^2\nonumber\\
&\quad+2\operatorname{Re}\langle B_x\psi_1,B_x\psi_2\rangle .
\label{eq:I-intro-interference}
\end{align}
The cross term distinguishes coherent alternatives from mutually exclusive recorded events, whose probabilities add. This equation is a statement about amplitudes. Identifying its left-hand side, after the appropriate normalization, with a detection chance is a separate statistical step.

For the double slit, Eq.~\eqref{eq:I-intro-interference} contains the pre-registration interference term.  A final position detector supplies registered and retained sectors \(Q_x\), where \(x\) labels a retained position bin, with effects \(E_x=V^\dagger Q_xV\). Under the statistical conditions,
\begin{equation}
p(x)=\langle\psi,E_x\psi\rangle
\end{equation}
gives the probability of the recorded position bin.  Coherent path alternatives therefore cannot be treated as already exclusive records whose positive weights are simply added.

The qubit of Eq.~\eqref{eq:II-false-record-state} already showed why choosing a future measurement basis does not create a present record. A physical interaction is required.
\label{eq:I-false-record-plus}
For a linear interaction $V$, such a correlation can take the form
\begin{equation}
 V|\psi\rangle=c_0|0\rangle|D_0\rangle+
                 c_1|1\rangle|D_1\rangle .
 \label{eq:I-cat-state}
\end{equation}
The two memory states describe possible readouts. The coherent sum does not itself select one of them. This is the familiar difficulty illustrated by Schr\"odinger's cat~\cite{Schrodinger1935}. Interference and quantum-eraser experiments further show why a reversible path marker and a retained detector record must be distinguished~\cite{Wheeler1978,ScullyDruhl1982,Englert1996,SchwindtKwiatEnglert1999}.

For the cat experiment, Eq.~\eqref{eq:I-cat-state} may remain a coherent vector of an enlarged source representation.  Orthogonal apparatus states alone are marker sectors, not a proof of a completed record.  Once an actual registration creates readable record sectors and the retention theorem applies, all later admitted readable observables are block diagonal across those sectors.  The retained algebra describes the classical alternatives available to the final reader. The assertion that one of them occurs is still the occurrence premise of Sec.~\ref{sec:actuality}.

For Wigner's friend~\cite{Wigner1961}, two descriptions can be consistent when they refer to two differently specified source processes.  The friend's record can be registered and retained relative to one continuation.  If a larger apparatus physically reverses that registration before it becomes a retained record of the enlarged experiment, the larger experiment has a later record boundary.  Mathematical unitary inversion alone does not show that such a reversal is physically constructible. This restricted claim does not assign one common set of absolute earlier outcomes to all possible coherent reversals. Such assignments face the Local Friendliness constraints~\cite{Bong2020}. If a comparison allows a reversal that destroys the friend's record, that record is outside the retained-event claim for that comparison. The restriction limits the claim about occurrence. It does not resolve every version of the friend problem.

For delayed choice and quantum erasure, a later controller selects the complete source process and its final registered records.  A marker correlation that is later coherently recombined was not thereby an earlier retained classical fact.  A registered record can lose readability through erasure, noise, loss of its carrier or a coherent reversal. Its retention is therefore a property of the specified continuation. A coherent reversal is a stronger operation than simply destroying the stored information.  No retrocausal change of a completed past record is required.

\subsection{EPR: conditional states, locality and predictive completeness}
\label{sec:I-2}

EPR asked whether the quantum description contains a counterpart for every element of physical reality.  Their sufficient reality criterion combined certain prediction with absence of disturbance~\cite{EPR}.  We address a narrower question: whether a specified source and apparatus determine every admitted record prediction, including predictions conditioned on a distant record.  A statement about the distant system's real condition is stronger than invariance of its unread local statistics.

The calculation presupposes a physically local operation.  For a constrained field theory, use the bounded distant observable algebra instead: a unitary source coupling $U_A$ satisfying $[U_A,B]=0$ for every distant $B$ gives $U_A^\dagger B U_A=B$.  This identity must follow from the actual coupling, not from norm preservation alone.  Conditioning can change correlations while the unread distant algebra remains unchanged.  No-signalling is not the stronger Bell factorization conditioned on hidden data~\cite{FewsterVerch2020,Bell}.

Let the joint represented predictive state be $\rho_{12}$. Before a record from the first system is supplied, the state relevant to all local predictions on the second system is
\begin{equation}
\rho_2=\operatorname{Tr}_1\rho_{12}.
\label{eq:rho2}
\end{equation}

For clarity, first take a source-derived local instrument with one Kraus map \(M_{a|x}\) for each readable outcome $a$ at setting $x$.  A general outcome can have several unread Kraus indices: replace \(M_{a|x}XM_{a|x}^\dagger\) below by \(\sum_\alpha M_{a\alpha|x}XM_{a\alpha|x}^\dagger\), and sum over both indices in the completeness relation.  The same partial-trace argument applies.  In the displayed one-map case,
\begin{equation}
\sum_a M_{a|x}^{\dagger}M_{a|x}=I.
\label{eq:kraus-complete}
\end{equation}
The Born theorem already proved above gives
\begin{equation}
p(a|x)=\operatorname{Tr}\!\left[(M_{a|x}^{\dagger}M_{a|x}\otimes I)\rho_{12}\right].
\label{eq:I-epr-outcome-prob}
\end{equation}
For \(p(a|x)>0\), if record $a$ is known, the state used for later predictions on system 2 is
\begin{equation}
\rho_{2|a,x}
=
\frac{
\operatorname{Tr}_1[
(M_{a|x}\otimes I)\rho_{12}
(M_{a|x}^{\dagger}\otimes I)]
}{
p(a|x)
}.
\label{eq:conditional-state}
\end{equation}
Different choices of $x$ may therefore give different conditional states. This is the circumstance emphasized by EPR~\cite{EPR}.

If the result at system 1 is not read, however, Eq.~\eqref{eq:kraus-complete} gives (with zero-probability conditional states irrelevant)
\begin{equation}
\sum_a p(a|x)\rho_{2|a,x}=\rho_2.
\label{eq:no-signalling}
\end{equation}
Thus a distant choice changes the conditional description appropriate after a particular distant record is supplied. It does not change the unread local state or its local statistics.

\paragraph{Selected normalization and a deterministic law.}
For a nonzero finite-dimensional CP map \(\Phi\), put \(R=\Phi^*(I)\) and \(F(\rho)=\Phi(\rho)/\operatorname{Tr}(R\rho)\) on its positive-success domain.  For \(a=\operatorname{Tr}(R\rho)>0\), \(b=\operatorname{Tr}(R\sigma)>0\), direct subtraction gives
\begin{align}
&F(t\rho+(1-t)\sigma)-tF(\rho)-(1-t)F(\sigma)\nonumber\\
&\quad=\frac{t(1-t)(a-b)}{ta+(1-t)b}
 [F(\rho)-F(\sigma)].
\label{eq:I-normalization-affinity-defect}
\end{align}
Consequently \(F\) is affine on this full domain precisely when either \(R=cI\), \(c>0\), or its output is constant: \(\Phi(X)=\operatorname{Tr}(RX)\tau\) for a fixed density operator \(\tau\).  To see necessity, if two inputs have unequal success traces, affinity forces their outputs to agree. Comparison with this pair forces the same output for every other input.  Linearity extends the resulting identity from the interior to all matrices.  The constant-output case is therefore part of the classification.

If one instead postulates the deterministic composite rule
\begin{equation}
\rho_{AB}\longmapsto
\frac{(\operatorname{id}\otimes\Phi)(\rho_{AB})}
{\operatorname{Tr}[(I\otimes R)\rho_{AB}]},
\label{eq:I-forced-normalization}
\end{equation}
a maximally entangled input gives Alice's marginal \(R^T/\operatorname{Tr}R\).  No change of her unread marginal for this rule requires \(R=cI\), including in the constant-output case.  This conclusion uses the composite and trace-statistical laws already established here. It cannot be used as an independent derivation of Born's rule.

For an explicit heralded process on a Bell pair, Bob's success and failure matrices
\begin{equation}
K=\operatorname{diag}(1,1/2),\qquad
L=\operatorname{diag}(0,\sqrt3/2)
\label{eq:I-heralded-filter}
\end{equation}
satisfy \(K^\dagger K+L^\dagger L=I\). Their unnormalized joint branches are
\[
\begin{aligned}
 (I\otimes K)|\Phi^+\rangle&=\frac{|00\rangle+\tfrac12|11\rangle}{\sqrt2},\\
 (I\otimes L)|\Phi^+\rangle&=\sqrt{\frac38}|11\rangle.
\end{aligned}
\]
  Success has probability \(5/8\) and gives Alice \(p(0|s)=4/5\). Failure has probability \(3/8\) and gives \(p(0|f)=0\).  Her unread probability is therefore \((5/8)(4/5)=1/2\).  The conditional shift from \(1/2\) to \(4/5\) is not signalling.  Forcing the success normalization deterministically would be a different source law.

The same Bell pair also shows how different distant settings give different conditional ensembles. In the $Z$ basis it is $(|00\rangle+|11\rangle)/\sqrt2$. With $|\pm\rangle=(|0\rangle\pm|1\rangle)/\sqrt2$, it is $(|++\rangle+|--\rangle)/\sqrt2$. Nevertheless,
\begin{equation}
\begin{split}
 \tfrac12(|0\rangle\langle0|+|1\rangle\langle1|)
 &=\tfrac12(|+\rangle\langle+|+|-\rangle\langle-|)\\
 &=I/2.
\end{split}
\label{eq:two-decompositions}
\end{equation}
The conditional descriptions differ while the unread local state stays the same.

Bohr's reply to EPR stressed that the conditions defining a physical prediction cannot in general be separated from the complete experimental arrangement~\cite{Bohr}. The present construction makes this part explicit. The preparation, apparatus, physical boundary and resolution are included in the specified source process. The conclusion, however, is not that reality is created by an observer. A completed detector has an objective record. What depends on the arrangement is which physical question was actually asked.

For a normalized vector $\psi$ and an actually specified effect, certainty has an exact meaning:
\[
 \langle\psi,E\psi\rangle=1
 \quad\Longleftrightarrow\quad E\psi=\psi,
 \qquad 0\le E\le I.
\]
This follows by applying $(I-E)^{1/2}$ to $\psi$. It does not turn conditional certainties in incompatible arrangements into simultaneous sharp values. Predictive completeness of the stated record law therefore neither establishes nor refutes EPR's stronger criterion about every element of reality.

\subsection{Predictive completeness and algebraic closure}
\label{sec:I-predictive-completeness}
We can now state the completeness claim. Under the stated source, detector and statistical conditions, every admitted finite record experiment has a fixed joint law. Grouping gives its marginals, and conditioning on positive-probability records gives its conditional predictions. For a finite terminal record family,
\begin{equation}
 F_{\rm rec}=\sum_r f_rQ_r,\qquad
 \langle F_{\rm rec}\rangle=\sum_r f_rp_r.
 \label{eq:I-all-record-means-example}
\end{equation}
Thus every statistical question that is a function of the completed record is determined. This is the precise sense of \emph{complete algebraic closure} used here. It requires no further probability assignment or additional collapse dynamics for those predictions. Infinite histories require the extension conditions of Sec.~\ref{sec:I-histories}.

A newly readable variable $\lambda$ requires an enlarged experiment that registers $(h,\lambda)$. If the same hypotheses hold there, $p(h,\lambda)=\operatorname{Tr}[\rho E(h,\lambda)]$. The theorem does not establish physical coverage of every conceivable experiment or uniqueness of an unreadable ontology.

\emph{Informational completeness} asks a different question. Does the available set of effects $\mathfrak M$ distinguish the represented states? Define the labelled map
\begin{equation}
 \Phi_{\mathfrak M}(\rho):E\longmapsto\operatorname{Tr}(\rho E),
 \qquad E\in\mathfrak M.
 \label{eq:I-informational-completeness-map}
\end{equation}
It is informationally complete on a stated class when this map is injective. A qubit tester measuring only $Z$ is predictively closed for that question, yet gives identical statistics for $|+\rangle$ and $|-\rangle$.

\emph{Finite-sample estimation} uses counts rather than exact probabilities. Even an informationally complete tester can yield different estimates under different estimation criteria. Hence
\begin{equation}
\begin{gathered}
 \text{predictive closure}\ne\text{tomographic identifiability}\\
 \ne\text{finite-sample estimator uniqueness}.
\end{gathered}
\label{eq:I-three-completenesses}
\end{equation}
The same distinction applies to processes: differences invisible to every available test remain undetermined, subject to positivity, causal normalization and the other physical conditions.

Can the hypotheses be established for every physically realizable experiment in a field sector? A readable variable whose statistics fall outside every admitted enlarged record law would challenge that proposed physical coverage. It would not refute a conditional theorem whose hypotheses fail for that experiment.

\section{Dynamics, interpretation and the scope of completeness}
\label{new:dynamics-main}
\subsection{What a field source must supply}
\label{sec:II-remaining-dynamics}
The familiar examples share the same probability structure. Does that common structure also determine their dynamics? Different physical field laws can realize the same reconstruction structure. Their domains, constraints, finite resolution and couplings remain supplied source data. A finite expression for a Euclidean functional is not yet a positive physical pairing for preparations, a real-time process or a material detector. Those identifications concern the same specified source. A Hermitian matrix chosen to implement a desired detector gate supplies a reference model, not a derivation of that model from an independently given field law.

The source conservation identity~\eqref{eq:source-transport}, writing and readback bounds, and history matching identity~\eqref{new:history-matching} test these independently specified physical constructions.

For interacting finite-proper-time QED the positive physical pairing, complete Lorentzian boundary process and material detector identification require source-specific proofs. A fixed-order cutting identity alone does not establish exact positivity at finite coupling. A finite-resolution Yang--Mills construction likewise does not by itself establish a relativistic four-dimensional continuum theory or a uniform positive mass gap. The present probability theorem depends on neither of those unproved sector claims. Its purpose is to make the common reconstruction and the physical obligations precise.

\subsection{A dynamical law for individual histories}
The two instruments in Sec.~\ref{sec:effect-instrument} have the same
unread evolution and different recorded laws. This leaves a further
question: what physical dynamics supplies an individual history?
A stochastic model can supply one realized trajectory by declaring a physical variable, its law, its coupling and its interpretation as a detector history. If a bounded population $p_r(t)$ is a martingale and tends almost surely to $1_{\{R=r\}}$, bounded convergence gives
\[
 \Pr(R=r)=\mathbb E[p_r(\infty)]=p_r(0).
\]
This is a valid conditional dynamical route to terminal Born weights when the initial populations equal the geometric weights. It depends on the stipulated stochastic law and convergence. It does not derive that law from the reconstructed Hilbert space. An auxiliary Gaussian integral or a reduced density matrix is not, by itself, an actual detector signal.

\subsection{What does emergence say about reality?}
The framework places the source before its representation. This does not make a reconstructed particle, field or detector record fictitious. It distinguishes the physical roles fixed by the source from a particular mathematical coordinate description of those roles. A particle basis, a wavefunction and a Hamiltonian describe the represented dynamics after the relevant construction. A symmetry and its central extension can fix the form of a canonical commutator. The numerical value and universality of $\hbar$ are not derived from the abstract joining rules.

A related distinction appears in Einstein's account of gravitation: a principle constraining the form of a physical law does not by itself choose its field equations~\cite{Einstein1950}. Spacetime requires the same care here. In the sector descriptions, causal order, boundary data or metric-dependent source operators are supplied. The reconstruction theorem does not independently derive Lorentzian signature or Einstein's field equations. Constructive approaches that identify geometry through physical signals, rods and clocks provide a useful comparison~\cite{AdlamLinnemannRead2025,ReadMenonFunctionalism}. Agreement between a mathematical symmetry and the geometry measured by matter is an additional physical result.

Penrose asks for more than a law assigning conditional probabilities. His gravitational argument questions the applicability of linear quantum evolution to superpositions with significantly different gravitational configurations~\cite{Penrose}. Calling the Hilbert space emergent does not settle that question. A specified gravitational source must establish its physical evolution, clocks and records. Conversely, his proposed lifetime scale does not itself determine a unique collapse law or its Born probabilities. These are dynamical questions left open by the common representation framework.

Englert's distinction between formalism and its interpretation is equally relevant~\cite{Englert2013}. Conditioning on a known record---state reduction in his terminology---requires no extra collapse dynamics. Taking actual occurrence as a physical starting point is a coherent position, but it is different from deriving its microscopic origin. The present paper states which position each theorem uses. Predictive completeness of an admitted class of experiments is not a theorem that the underlying ontology is unique or that no new source dynamics can exist.

\section{Conclusion}
\label{joint:conclusion}
\label{sec:II-22}
\label{sec:I-conclusion}
The source reconstruction gives the source-to-record construction of Fig.~\ref{fig:source-record-probability}. The primitive finite resolutions and domain data enter before the quantum representation and remain distinct from auxiliary numerical refinement. A particular field theory supplies a realization of the source law $Z$.

What follows under the representation conditions is the internal mathematical language used for the admitted quantum predictions: amplitudes, Hilbert geometry, represented operators and processes, record algebras and effects. Positivity of the physical pairing is essential. Conservation is a separate dynamical property, and complete nonunitary sources require their own normalization conditions. Complex numbers alone do not establish nonclassicality.

The probability analysis determines Born's formula from preparation sufficiency, independent fair mixing and sharp reader calibration on the stated domain. The local proof avoids a probability postulate expressed as a universal function of squared branch length. A second route obtains the pure retained-record law from the explicit completeness, locality, calibration and auxiliary-control conditions of Sec.~\ref{sec:causality-route}. It retains an explicit statistical input about physical preparations. The finite history criterion and matching theorem show exactly how an independently supplied source history law can establish or fail that input. They do not yet provide that law for every interacting field source.

Three distinctions remain essential. Registration requires an actual interaction that writes a readable record. Retention concerns its persistence under specified continuations. Occurrence concerns the one record realized in a completed run. Decoherence can support retention, while conditioning on a known outcome gives subsequent predictions without a second collapse dynamics. Neither operation by itself chooses that outcome.

The complete algebraic closure claimed here is consequently exact within the stated conditions and admitted record domain. The source and apparatus determine all finite joint and marginal laws of their completed records, together with conditional laws for records of positive probability. A newly readable variable requires an enlarged experiment with the same hypotheses checked again.

The framework therefore answers one foundational question while leaving a deeper one open. It shows how the familiar mathematics of quantum prediction can be reconstructed from a more primitive physical architecture under explicit conditions. It does not establish that all source laws, spacetime or mathematics itself emerge from that architecture. Which physical law supplies the pairing, dynamics and statistical conditions, and what accounts for an individual record, remain questions for a concrete source theory.

\appendix
\section{Comparison and splitting as a separate characterization}
\label{app:comparison}
A separate characterization compares events in different complete experiments. It is useful for locating the extra statistical step that geometry alone cannot provide. Fix a normalized preparation $\psi$. For an event $S$ in experiment $C$, define its geometrical value
\[
 a_C(S)=\langle\psi,E_C(S)\psi\rangle.
\]
This number is not yet assigned a probability interpretation.

\paragraph{Complete-event comparison condition.}
Suppose every covered event can be compared with a reference success flag of the same geometrical value $a$. Require those physical comparisons to preserve probabilities. Require also that the reference probability be independent of all other retained records and correlations at fixed $a$. The condition covers elementary outputs, their grouped events, and the empty and sure events. These requirements give one function $f$ with
\[
 p_C(S)=f(a_C(S)).
\]
They are statistical requirements on complete experiments. A unitary relating two branch vectors does not prove them.

Suppose the apparatus supplies scalar splitters with every required setting, so that a three-output experiment can have squared branch lengths $s,t,1-s-t$. The first two records and their union are all covered by the comparison condition. Their probabilities are $f(s),f(t)$ and $f(s+t)$. Ordinary addition within that experiment gives
\begin{equation}
 f(s+t)=f(s)+f(t),\qquad s,t\ge0,\quad s+t\le1.
 \label{eq:I-Cauchy-additivity}
\end{equation}
Positivity makes $f$ monotone. Normalization gives $f(1)=1$, and repeated addition gives $f(m/n)=m/n$. Rational lower and upper bounds then give $f(a)=a$ for all $a\in[0,1]$. No separate continuity assumption is required. Conversely, Born's formula satisfies those statistical comparisons wherever the stated geometric identifications hold. This route is a conditional characterization, not an independent source derivation of the comparison condition.

\paragraph{Spin-1: coordinates and different apparatus arrangements.}
Prepare
\[
 |\psi\rangle=\tfrac12|+\rangle+\tfrac1{\sqrt2}|0\rangle+\tfrac12|-\rangle.
\]
One arrangement distinguishes $+,0,-$. Another distinguishes $(|+\rangle+|-\rangle)/\sqrt2$, $|0\rangle$ and $(|+\rangle-|-\rangle)/\sqrt2$. Both are three-dimensional projective arrangements, if physically available. The second is not in general the measurement of a spin component along a rotated spatial axis. Their squared branch lengths are $(1/4,1/2,1/4)$ and $(1/2,1/2,0)$. The selected projector $P_0$ is unchanged.

The hypothetical elementary-record rule
\begin{equation}
 p_j=\frac{a_j^2}{\sum_k a_k^2}
 \label{eq:I-squared-weight-counterrule}
\end{equation}
gives selected probabilities $2/3$ and $1/2$. Define probabilities of unions by addition. Each arrangement then has positive normalized additive probabilities, but the two probabilities for $P_0$ differ. A passive basis change, which transforms the state and all projectors together, would leave each physical experiment unchanged. Changing the apparatus is a different operation.

For lengths $(1/4,1/4,1/2)$, the same rule gives $(1/6,1/6,2/3)$. The union of the first two records has squared length $1/2$ and probability $1/3$. An equal-weight binary reference assigns $1/2$. This shows why the comparison must cover grouped events. Grouping recorded orthogonal alternatives adds their squared norms. It is not the coherent superposition of two arbitrary amplitudes.

\paragraph{Retaining side information.}
A reversible flag-writing control can act as
\[
 |b,r\rangle\longmapsto|b\mathbin\oplus1_S(r),r\rangle.
\]
It records whether $r\in S$ without deleting the old label. The flag effect on a ready input is $Q_S$. This operator identity does not show that the flag probability is independent of the retained label or its registration history. The statistical condition above says exactly that for the covered comparisons. The preparation theorem in the main text takes a different set of premises and derives the corresponding probability equality on its own domain.

\section{Stability of the probability law}
\label{new:scalar-stability}
\paragraph{Stability of scalar refinement.}
\label{thm:I-approximate-refinement-stability}
This theorem assumes one error bound for every pair of scalar alternatives. A finite calibration sample does not itself establish that uniform bound, and it is not a sampling confidence interval.

Let $\delta\ge0$ and let $f:[0,1]\to\mathbb R$ satisfy
$f(0)=0$, $f(1)=1$, $f(x)\ge0$, and
\[
 |f(s+t)-f(s)-f(t)|\le\delta
 \qquad(s,t\ge0,\ s+t\le1).
\]
Then, without any continuity or measurability assumption,
\[
 \sup_{0\le x\le1}|f(x)-x|\le2\delta.
\]
In particular, for every integer $n\ge1$,
\[
 |f(x)-x|\le\frac1n+(2n+1)\delta.
\]
The constant $2$ is a sufficient bound. Optimality is not claimed.

\emph{Proof.}
Taking $t=1-s$ and using nonnegativity gives
$0\le f(s)\le1+\delta$.
For $u\ge0$ write $u=\lfloor u\rfloor+\{u\}$ and define
\[
 F(u)=\lfloor u\rfloor+f(\{u\}).
\]
We first show $|F(u+v)-F(u)-F(v)|\le2\delta$. Write $s=\{u\}$ and $t=\{v\}$. If $s+t<1$, the defect is $f(s+t)-f(s)-f(t)$ and is bounded by $\delta$. If $s+t\ge1$, set $a=1-s$ and $r=s+t-1$, so that $s+a=1$ and $a+r=t$. The defect is
\[
\begin{aligned}
 &1+f(r)-f(s)-f(t)\\
 &\quad=\bigl[1-f(s)-f(a)\bigr]
  +\bigl[f(a)+f(r)-f(t)\bigr],
\end{aligned}
\]
whose absolute value is at most $2\delta$. The doubling case therefore gives
\[
 \left|2^{-j-1}F(2^{j+1}x)-2^{-j}F(2^jx)\right|
 \le2\delta\,2^{-j-1}.
\]
Summing for $0\le j<N$ yields
\[
 |2^{-N}F(2^Nx)-F(x)|\le2\delta(1-2^{-N}).
\]
Since $|F(u)-u|=|f(\{u\})-\{u\}|\le1+\delta$, we have $2^{-N}F(2^Nx)\to x$. Thus $|F(x)-x|\le2\delta$. For $0\le x<1$ we have $F(x)=f(x)$, and at $x=1$ both equal one. The final bound follows because $n\ge1$ and $\delta\ge0$ imply $2\delta\le n^{-1}+(2n+1)\delta$.
\hfill$\Box$

\paragraph{Operational meaning of the stability hypothesis.}
The error $\delta$ is a uniform defect in a single scalar probability law on the entire refinement triangle. Obtaining that law approximately from physical branch controls is an additional operational premise. Finite sampling or a finite menu of calibrated splitters does not alone establish this uniform hypothesis. If a measured or modelled branch probability $p(x)$ obeys $|p(x)-f(x)|\le\varepsilon$, the theorem gives $|p(x)-x|\le\varepsilon+2\delta$. Here $\varepsilon$ and $\delta$ are bounds on probability differences in the stated apparatus comparisons. They are not automatically equal to an isometry error $\eta$ or a Gram-effect error $2\eta$. An amplitude-to-probability error estimate must be supplied independently before it is used in a derivation of the probability law. Applying the already derived Born formula is appropriate only for calibration performed after the probability theorem has been established.

\section{Statistical comparisons and calibration counterexamples}
\label{sec:I-probability-independence}

The comparison-and-splitting characterization is stated in Appendix~\ref{app:comparison}. The main preparation-based theorem is stated in Sec.~\ref{sec:I-born-main}. The examples here test the comparison premise and distinguish finite calibration from an exact statement over a full family of experiments.

\paragraph{Counterexample without complete-event equivalence.}
Retain the actual finite registration tree as part of the source context.  At each node assign probabilities \(q_j^2/\sum_kq_k^2\) to geometric branch weights \(q_j\), and use the normalized selected branch vector for subsequent geometric evolution.  Every transition kernel normalizes. Multiplication defines a positive additive law on all finite histories.  Summing any appended subtree gives one, so later refinements preserve every earlier marginal.  Null branches have zero probability, certain branches have probability one, and ordinary isometric splitters and post-record unitary controls remain available.  Direct three-way registration with branch maps
\begin{equation}
(I/\sqrt2,I/2,I/2)
\label{eq:I-context-counterexample-maps}
\end{equation}
gives \((2/3,1/6,1/6)\).  First registering two equal branches and then equally refining only the second gives the same final branch maps but probabilities \((1/2,1/4,1/4)\).  Their registration histories differ.  The law violates the complete-event comparison condition while retaining within-context additivity and temporal refinement consistency.  It is not asserted to satisfy ensemble affinity or unrestricted relativistic composition.  Those would be additional conditions, not consequences of branch-local controls.  Thus the operator refinement identity cannot alone supply the missing statistical equivalence.

\paragraph{A countable calibration alternative.}
An all-real family of reference settings is sufficient but can be replaced by a different explicit comparison premise.  Suppose that for every \(n\ge0\) an admitted calibration experiment has \(2^n\) equal-Gram leaves connected by probability-preserving complete-event equivalences.  Normalization and finite additivity give probability \(m2^{-n}\) to a union of \(m\) leaves.  For a target event of geometric weight \(a\), put
\begin{equation}
l_n=2^{-n}\lfloor2^na\rfloor,\qquad
u_n=\min(1,l_n+2^{-n}).
\label{eq:I-dyadic-calibration}
\end{equation}
Require complete comparison experiments in which an event equivalent to the \(l_n\) calibration is contained in an event equivalent to the target, and an event equivalent to the target is contained in an event equivalent to the \(u_n\) calibration.  Include endpoints and all coarse unions used.  Positivity and additivity then imply \(l_n\le p(E)\le u_n\) for every \(n\), hence \(p(E)=a\), without statistical continuity.  This removes the continuum of control settings only by adding actual order-comparison witnesses. It does not derive those witnesses from geometry.  A finite hierarchy gives only a finite bracket.  Dense calibrated settings alone are insufficient: a disconnected binary context of weights \((3/4,1/4)\) can follow the squared-weight rule and assign \(9/10\), although every uniform dyadic context is calibrated correctly.

\paragraph{What finitely many calibrations establish.}
Suppose actual comparison experiments supply event inclusions $L\subseteq E\subseteq U$, or equivalent comparisons with the target whose probability transfers have been independently justified. Suppose the certified probability bounds are
\[
\begin{gathered}
 p(L)\ge l-\varepsilon,\qquad p(U)\le u+\varepsilon,\\
 0\le l\le a\le u\le1.
\end{gathered}
\]
Here $a$ is the geometrical value assigned to the target. Positivity and finite additivity on the comparison experiments imply
\[
 \boxed{\max(0,l-\varepsilon)\le p(E)
       \le\min(1,u+\varepsilon).}
\]
Consequently
\[
 |p(E)-a|\le\varepsilon+\max(a-l,u-a).
\]
A calibration mesh of maximum spacing $h$ gives the uniform sufficient bound $\varepsilon+h$. Multiple independently certified comparisons may be combined by taking the largest lower bound and smallest upper bound. For the stated order constraints these bounds are sharp: any consistent intermediate value extends to a monotone assignment on the finite ordered comparison set. Exact uniqueness requires comparisons at arbitrarily fine scales or an additional independently established restriction on the law. An operator inequality between effects does not by itself establish the corresponding probability comparison before the probability theorem.

\paragraph{Finite calibration is insufficient even for smooth laws.}
Let $S\subset(0,1)$ be a finite set invariant under $q\mapsto1-q$, with
$n=|S|$. Define
\[
\begin{aligned}
 g(x)&=x(1-x)(2x-1)\prod_{q\in S}(x-q)^2,\\
 h(x)&=x+\frac{g(x)}{2(2n+4)}.
\end{aligned}
\]
On $[0,1]$ each linear factor has magnitude at most one, and the product
rule gives $|g'(x)|\le2n+4$. Hence $h'(x)\ge1/2$.
The polynomial $h$ has $h(0)=0$, $h(1)=1$,
$h(1-x)=1-h(x)$ and $h(q)=q$ at every calibration setting, but it is
not the identity function. For $S=\{1/4,1/2,3/4\}$,
\[
 h(1/3)-1/3=-\frac5{40310784}.
\]
Thus finite calibration, smoothness, monotonicity and binary complement normalization do not force the Born assignment. The polynomial does not satisfy the full cross-experiment additive comparison premise of the Born theorem.

\section{Mixed preparations and a complete reader domain}
\label{app:full-reader}
The local disk proof applies to each pure target when its preparations are physically available. A complete process with a mixed terminal preparation can instead use the following extension. Suppose a fixed sharp reader $Q$ has a probability $q(\rho)$ on all positive trace-one trace-class operators on its declared Hilbert space. Require preparation sufficiency, independent fair mixing, $0\le q\le1$, and calibration on every pure state in the range and kernel of $Q$. These are richer domain assumptions than one target disk.

The segment argument in the main proof gives affinity for every real mixture coefficient. For $A\ge0$ trace class define
\[
 \ell(A)=\begin{cases}
 (\operatorname{Tr}A)q(A/\operatorname{Tr}A),&A\ne0,\\
 0,&A=0.
 \end{cases}
\]
Affinity gives additivity on the positive cone, and $0\le\ell(A)\le\operatorname{Tr}A$. For a self-adjoint trace-class operator set $\ell(A-B)=\ell(A)-\ell(B)$ with $A,B\ge0$. This is well defined: two such decompositions give $A+B'=A'+B$, and cone additivity equates the differences. Extend complex linearly.

The resulting functional is bounded. For $T=U|T|$, the positive functional Cauchy--Schwarz inequality, obtained by expanding $\ell((A+zB)^*(A+zB))\ge0$ for Hilbert--Schmidt $A,B$, gives
\[
 |\ell(T)|^2\le\ell(|T|)\ell(U|T|U^*)\le\|T\|_1^2.
\]
Thus trace-class duality gives one bounded operator $F$ such that $\ell(T)=\operatorname{Tr}(TF)$. Positivity of $\ell$ and of $\operatorname{Tr}-\ell$ gives $0\le F\le I$. Calibration gives $Fx=x$ on the range of $Q$ and $Fx=0$ on its kernel. Indeed, zero expectation of a positive operator implies that its positive square root annihilates that vector. Consequently $F=Q$ and
\[
 q(\rho)=\operatorname{Tr}(\rho Q).
\]
This establishes Eq.~\eqref{eq:born-general-process} for a normal represented apparatus whose output preparations meet these hypotheses. A restricted family of physically produced mixtures also suffices when all components and mixing procedures used are admitted. The algebraic density-operator domain by itself does not construct those procedures.

\paragraph{A finite preparation error.}
Once the trace law is established, a preparation satisfying $\langle\psi,\sigma\psi\rangle\ge1-\epsilon$ obeys
\[
 |\operatorname{Tr}(\sigma E)-\langle\psi,E\psi\rangle|\le\sqrt\epsilon
 \qquad(0\le E\le I).
\]
Diagonalize $\sigma=\sum_jw_j|\phi_j\rangle\langle\phi_j|$. On the two-vector span, the difference of the pure projections has eigenvalues $\pm\sqrt{1-|\langle\psi,\phi_j\rangle|^2}$. Convexity of trace norm and Cauchy--Schwarz give
\[
 \tfrac12\|\sigma-P_\psi\|_1
 \le\sum_jw_j\sqrt{1-|\langle\psi,\phi_j\rangle|^2}
 \le\sqrt\epsilon.
\]
The expectation bound follows from trace duality and the zero trace of the difference. It is sharp for two pure states with overlap squared $1-\epsilon$. This is an error bound after the probability theorem, not a substitute for its statistical conditions.

\section{A finite test of preparation sufficiency and history matching}
\label{new:history-appendix}

An independently reconstructed state can predict a reader only if the physical information omitted from that state has no effect on the admitted records. This appendix gives a finite test of that property and a stronger condition identifying the predictions with independently reconstructed quantum instruments. The initial and transition laws are supplied physical data throughout. The results are conditional diagnostics and provide no construction of those laws for interacting finite-proper-time sources.

\subsection*{An independently specified history law}
Let $X$ be a measurable space of complete initial data and let $\nu_\alpha$ be the probability law supplied for preparation procedure $\alpha$. For a specified experiment $D$, let $k_D(B\mid x)$ be a probability kernel on a declared measurable record space. These data give
\begin{equation}
 p_D(B\mid\alpha)
 =\int_X k_D(B\mid x)\,\nu_\alpha(dx).
 \label{new:history-law}
\end{equation}
In a deterministic model with a measurable record map $R_D$, the kernel is $\mathbf1_B(R_D(x))$. In a stochastic model it is determined by the stipulated physical noise and evolution. Equation~\eqref{new:history-law} uses that independently specified law. It assigns no trace formula to the records.

Suppose the source also supplies a represented preparation $\rho_\alpha$. Its sufficiency for the declared experiments means
\begin{equation}
 \rho_\alpha=\rho_\beta
 \quad\Longrightarrow\quad
 \int_X k_D(B\mid x)\,(\nu_\alpha-\nu_\beta)(dx)=0
 \label{new:history-sufficiency}
\end{equation}
for every admitted $D,B$. The condition concerns the responses to the differences between preparation laws. Equality of the laws themselves is a stronger condition.

\subsection*{The finite response criterion}
Take a model with $N\ge1$ complete states, finitely many controls $c$ and finitely many records $r$ for each control. Assume that every finite control sequence is admitted. For $a=(c,r)$, let $T_a$ be a nonnegative real $N\times N$ matrix with
\begin{equation}
 \sum_r T_{c,r}\mathbf1=\mathbf1.
 \label{new:history-normalization}
\end{equation}
Rows label the current state and columns the next state. The conditional transition and record probabilities are the entries of $T_a$. Thus a probability column $\mu$ becomes the unnormalized column $T_a^T\mu$ after the specified record. All information that can affect a later record is included in these $N$ states.

For a word $h=(a_1,\ldots,a_n)$ define
\begin{equation}
\begin{gathered}
 w_h=T_{a_1}\cdots T_{a_n}\mathbf1,\\
 p_\mu(h)=\mu^T w_h,\qquad w_{\varnothing}=\mathbf1.
\end{gathered}
 \label{new:history-response}
\end{equation}
Let $S:\mathbb R^N\to\mathbb R^d$ be the independently supplied linear map to represented preparation coordinates. Assume that normalization is retained, so $\mathbf1\in\operatorname{Ran}S^T$. Set
\begin{equation}
 \mathcal O=\operatorname{span}\{w_h:|h|\ge0\},
 \qquad L=\operatorname{Ran}S^T.
 \label{new:history-observable-space}
\end{equation}
On the full initial probability simplex, the following criterion is exact:
\begin{equation}
 \begin{gathered}
 S\mu=S\nu\ \Longrightarrow\ p_\mu(h)=p_\nu(h)
 \quad\text{for every word }h
 \\
 \Longleftrightarrow\quad \mathcal O\subseteq L.
 \end{gathered}
 \label{new:history-finite-criterion}
\end{equation}

\paragraph{Proof.}
If $w_h=S^Tb_h$, then $\mu^Tw_h=b_h^TS\mu$, proving sufficiency. Conversely, let $v\in\ker S$. Normalization in the row space of $S$ gives $\mathbf1^Tv=0$. For a strictly positive probability column $\xi$ and sufficiently small $\epsilon>0$, both $\xi+\epsilon v$ and $\xi-\epsilon v$ are probability columns with the same represented coordinates. Their equal responses give $v^Tw_h=0$. Therefore $w_h\in(\ker S)^\perp=L$ for every word, proving necessity.

The criterion has a finite witness bound. Define
\begin{equation}
 \mathcal O_0=\operatorname{span}\{\mathbf1\},
 \qquad
 \mathcal O_{j+1}=\mathcal O_j+\sum_a T_a\mathcal O_j.
 \label{new:history-iteration}
\end{equation}
Induction identifies $\mathcal O_j$ with the span of words of length at most $j$. If $\mathcal O_{j+1}=\mathcal O_j$, this space is invariant under every $T_a$ and the iteration has finished. Each earlier step increases its dimension. Starting at dimension one in $\mathbb R^N$ therefore gives $\mathcal O_{N-1}=\mathcal O$. If sufficiency fails, at least one word of length at most $N-1$ has $w_h\notin L$ and distinguishes equal-state preparations by the preceding proof. This bound assumes a finite complete-state model. A finite detector alphabet alone supplies no bound on $N$.

For a restricted set of admitted preparations, the exact criterion instead requires each response to annihilate the span of its actual equal-state differences. The full-simplex criterion remains sufficient, while its necessity uses access to the perturbations in the proof.

\subsection*{An exact discrepancy certificate}
For a fixed word, define
\begin{equation}
 \Delta_h=\max_{\substack{\mu,\nu\ge0\,,\ \mathbf1^T\mu=\mathbf1^T\nu=1\\
                         S\mu=S\nu}}
                |w_h^T(\mu-\nu)|.
 \label{new:history-discrepancy}
\end{equation}
Then
\begin{equation}
 \boxed{\displaystyle
 \Delta_h=2\inf_{b\in\mathbb R^d}\|w_h-S^Tb\|_\infty.}
 \label{new:history-distance}
\end{equation}
Thus the best uniform approximation of the actual response by represented coordinates measures exactly the largest unresolved probability difference.

\paragraph{Proof.}
Every feasible difference $v=\mu-\nu$ satisfies $Sv=0$ and $\|v\|_1\le2$. Conversely, these two conditions imply $\mathbf1^Tv=0$. Write $v=v_+-v_-$, with each part having mass $q\le1$. For any probability column $\xi$, the columns
\[
 \mu=v_++(1-q)\xi,\qquad
 \nu=v_-+(1-q)\xi
\]
realize that difference. The feasible differences are therefore exactly the radius-two $1$-norm ball in $\ker S$.

Put $w=w_h$ and $d_0=\inf_{z\in L}\|w-z\|_\infty$. For $Sv=0$, the inequality
\[
 |v^Tw|=|v^T(w-z)|\le\|v\|_1\|w-z\|_\infty
\]
gives $\Delta_h\le2d_0$. For $d_0>0$, separate the linear space $L$ from the open cube $\{x:\|x-w\|_\infty<d_0\}$. Finite-dimensional separation gives a nonzero vector $v$ annihilating $L$ with $v^Tw\ge d_0\|v\|_1$. The preceding inequality gives equality. Rescale to $\|v\|_1=1$. Then $Sv=0$, and the feasible difference $2v$ attains $2d_0$. If $d_0=0$, the upper bound already proves the result. The closed finite-dimensional space $L$ has a best approximant, and every such approximant has a preimage under $S^T$.

At a fixed finite horizon, choose one control policy and let its possible transcripts be $h$. A policy may choose later controls from earlier records. Its responses satisfy $\sum_h w_h=\mathbf1$. For a grouped event $B$ of transcripts, put $w_B=\sum_{h\in B}w_h$. With total variation defined by the maximum event-probability difference,
\begin{equation}
 \max_{S\mu=S\nu}\operatorname{TV}(P_\mu,P_\nu)
 =\max_B\,2\inf_b\|w_B-S^Tb\|_\infty,
 \label{new:history-grouped}
\end{equation}
where the preparations in the maximum are probability columns. Indeed, total variation is $\max_B|(\mu-\nu)^Tw_B|$. The finite maxima over events and preparations can be interchanged, and Eq.~\eqref{new:history-distance} applies to each $B$. A bound for each individual word gives the safe estimate $\operatorname{TV}(P_\mu,P_\nu)\le\frac12\sum_h\Delta_h$. A common small bound on single words can accumulate across a grouped event.

\subsection*{Matching the represented instrument}
Sufficiency identifies a response as a function of represented coordinates. Identifying that function with a particular reconstructed reader also requires its record labels and dynamics to agree. Let $\ell^T$ be the trace functional on the represented coordinates, with
\begin{equation}
 \ell^TS=\mathbf1^T.
 \label{new:history-trace}
\end{equation}
For each recorded operation $a$, let $A_a$ be its independently constructed linear state map. Assume the same labelled operations obey
\begin{equation}
 \boxed{ST_a^T=A_aS.}
 \label{new:history-matching}
\end{equation}
Then every initial probability column and every word satisfy
\begin{equation}
 p_\mu(h)=\ell^TA_{a_n}\cdots A_{a_1}S\mu.
 \label{new:history-matched-probability}
\end{equation}

\paragraph{Proof.}
Successive use of Eq.~\eqref{new:history-matching} gives
\[
 ST_{a_n}^T\cdots T_{a_1}^T\mu
 =A_{a_n}\cdots A_{a_1}S\mu.
\]
Apply $\ell^T$ and use Eq.~\eqref{new:history-trace}. The left side becomes $\mathbf1^TT_{a_n}^T\cdots T_{a_1}^T\mu=\mu^Tw_h$. The same proof works on a restricted preparation domain when the matching identity holds on the linear span of all its reachable unnormalized columns.

If $S\mu$ represents a density operator $\rho_\mu$ and $A_a$ represents an independently constructed completely positive instrument map $\mathcal I_a$, this equality reads
\begin{equation}
 \begin{aligned}
 p_\mu(h)
 &=\operatorname{Tr}\bigl[\mathcal I_h(\rho_\mu)\bigr]
 =\operatorname{Tr}(\rho_\mu G_h),\\
 \mathcal I_h&=\mathcal I_{a_n}\circ\cdots\circ\mathcal I_{a_1},
 \qquad G_h=\mathcal I_h^*(I).
 \end{aligned}
 \label{new:history-instrument-trace}
\end{equation}
Here positivity of the represented states, complete positivity of the maps and trace preservation after summing each instrument's records are separate hypotheses. The left side follows from the supplied history law. The equality with the trace expression follows from matching. Defining the transition law by that desired trace expression would assume the equality being tested. Constructing state coordinates from the full response table would establish a different operational representation. The diagnostic here compares the independently supplied objects on their declared preparation domain.

\subsection*{A delayed-memory counterexample}
Consider the declared deterministic process with complete state $(s,m)\in\{0,1\}^2$. Each step records the visible bit $s$ and then swaps $s$ and $m$. Its actual matrices are
\begin{equation}
 T_r\bigl((s,m),(s',m')\bigr)
 =\mathbf1_{\{s=r\}}\mathbf1_{\{(s',m')=(m,s)\}}.
 \label{new:history-memory-model}
\end{equation}
Order the states as $00,01,10,11$ and let $S$ retain only the visible marginal,
\[
 S\mu=(\mu_{00}+\mu_{01},\ \mu_{10}+\mu_{11})^T.
\]
The point preparations at $00$ and $01$ have identical represented coordinates and both give first record $0$. Their second records are respectively $0$ and $1$. Hence the word $00$ has probabilities one and zero. Its response is $(1,0,0,0)^T$, while $L$ consists of vectors $(a,a,b,b)^T$. Its distance from $L$ is $1/2$, attained at $(1/2,1/2,0,0)^T$. Equation~\eqref{new:history-distance} therefore gives $\Delta_{00}=1$. The first-record test passes and the history test detects the returning memory. This supplied finite model illustrates the diagnostic. Establishing the corresponding history law, representation and matching for a physical FPT source remains a separate construction.

\paragraph{Scope of a finite positive representation.}
If $S$ maps the full probability simplex affinely to density operators, its image is the convex hull of the $N$ states represented by the point preparations. A pure state in this image must equal every state with nonzero weight in its convex decomposition. Indeed, positivity forces each such state's support to lie in the same one-dimensional range. The image therefore contains at most $N$ distinct pure states and cannot cover all pure qubit preparations. The finite criterion applies to its stated preparation domain. A source with continuous or infinitely many complete states requires its own extension of the argument.

\section{History consistency and repeated frequencies}
\label{app:history-proof}
For a finite sequence of normalized instruments, positivity gives $p(h)=\operatorname{Tr}\mathcal I_h(\rho)\ge0$. Sum the last record first. The dual of the summed last instrument sends $I$ to $I$, so the resulting sum is the probability of the preceding prefix. Repeating proves $\sum_h p(h)=1$ and every prefix identity. Adaptive controls cause no change: at each already specified prefix, the chosen complete instrument is normalized.

If two recording operations are causally incomparable and their full selective maps commute, exchanging them leaves every $\mathcal I_h$ unchanged. Any two linear listings of a finite partial order can be related by adjacent exchanges of incomparable entries. To prove this, move the first entry of the target listing leftward in the other listing. Every entry it passes is incomparable, since both listings respect the order. Delete that entry and repeat. This proves listing independence without inferring spacelike commutation from covariance.

Marginalization concerns the same experiment. Summing a stored record is different from removing the coupling that produced it. For example, an unread sharp $Z$ measurement sends a qubit $|+\rangle\langle+|$ to $I/2$, changing the probability of a later $X=+1$ record from one to one half. Deleting the first measurement would leave probability one. A history description that omits relevant memory can also fail, as Appendix~\ref{new:history-appendix} demonstrates.

\paragraph{Error accumulation over a finite history.}
Fix a control policy. Let the actual conditional law at prefix $h_{<k}$ differ from a reference law by total variation at most $\delta_k(h_{<k})$. Both laws must be defined at the prefixes being compared. For an $n$-record history,
\[
 \operatorname{TV}(P,Q)\le
 \min\!\left\{1,\sum_{k=1}^n
 \mathbb E_P[\delta_k(H_{<k})]\right\}.
\]
Insert intermediate laws using the first $k$ actual transitions and the remaining reference transitions. At the $k$th replacement their common prefix law is the actual one. Applying a common continuation cannot increase total variation, so the difference is bounded by the displayed expectation. Summing proves the result. Single-event error control therefore extends to finite histories with an explicit accumulation bound, provided it holds on the relevant complete prefixes.

\paragraph{The concentration calculation.}
For independently reset represented preparations, squared norms of orthogonal record strings factor as $\prod_{j=1}^N a_{r_j}$, where $a_r\ge0$ and $\sum_r a_r=1$. For a fixed record $r$, summing strings at fixed count gives the binomial weights with parameter $a_r$. This is a mathematical identity of product norms.

Let $X$ be a binary variable with those algebraic weights and let $g(t)=\log\mathbb E e^{t(X-a_r)}$. Then $g(0)=g'(0)=0$ and $g''(t)$ is a tilted binary variance, at most $1/4$. Hence $g(t)\le t^2/8$. Product factorization and the exponential bound give
\[
\begin{aligned}
 \Pr\!\left(\frac1N\sum_jX_j-a_r\ge\delta\right)
 &\le\inf_{t>0}e^{-tN\delta+Nt^2/8}\\
 &=e^{-2N\delta^2}.
\end{aligned}
\]
The other tail has the same bound. A union bound over $K$ labels gives Eq.~\eqref{eq:I-frequency-concentration-example}. Before a statistical interpretation, the symbol $\Pr$ in this calculation denotes the normalized product weights used to sum squared norms. Identifying them with frequencies of actual records requires the single-run probability law and independent physical resets. Correlated preparations or a returning environment require the full history law instead.

\section{A finite writing interaction and its readout}
\label{app:probe}
This exact bounded model illustrates the apparatus conditions. It is not a derivation of a material detector from interacting QED. Let $P_1,\ldots,P_K$ be orthogonal input projections summing to identity. A probe has ready state $|0\rangle$ and orthogonal labels $|r\rangle$. Set
\[
 G=\sum_r P_r\otimes(|0\rangle\langle r|+|r\rangle\langle0|),
 \qquad H=\frac{\hbar\theta}{\tau}G.
\]
Since $G^3=G$, the exact finite-time evolution is
\[
 U_\theta=I+(\cos\theta-1)G^2-i\sin\theta\,G.
\]
On the ready input it gives
\[
 U_\theta|\psi,0\rangle
 =\cos\theta|\psi,0\rangle
 -i\sin\theta\sum_rP_r|\psi\rangle|r\rangle.
\]
The selective maps before the final memory copy have Kraus operators $M_0=\cos\theta I$ and $M_r=-i\sin\theta P_r$. They obey $\sum_jM_j^\dagger M_j=I$. At $\theta=\pi/2$ the ready component vanishes.

At a general finite coupling, add a separate memory with ready state $|\bot\rangle$ orthogonal to all its written states $|j\rangle$. A probe-controlled swap implements
\[
 |j\rangle_P|\bot\rangle_C\longmapsto|j\rangle_P|j\rangle_C.
\]
The written inconclusive record $j=0$ is now distinct from readiness. The effects are $E_0=\cos^2\theta I$ and $E_r=\sin^2\theta P_r$. This copies the orthogonal probe basis. It does not distinguish the nonorthogonal conditional vectors $\cos\theta|0\rangle-i\sin\theta|r\rangle$ with certainty. Retention requires the continuation conditions of the main text.

A dispersive circuit-QED apparatus can motivate such an effective interaction~\cite{BakrBoundary2025}. Establishing that identification requires the material source, its controls and a bound on omitted dynamics. More generally, if $V$ and $V_0$ are two isometries with the same output effect $0\le Q\le I$, then
\[
 \|V^\dagger QV-V_0^\dagger QV_0\|
 \le2\|V-V_0\|.
\]
Expand the difference as $(V-V_0)^\dagger QV+V_0^\dagger Q(V-V_0)$. This proves the bound and, after the probability theorem, the corresponding uniform probability error. An independently derived generator and physical embedding $J(t)$ can test a proposed effective Hamiltonian through
\[
 H_Z(t)J(t)-i\hbar\dot J(t)-J(t)H_{\rm ref}(t).
\]
For unitary evolutions with the required differentiable domains, the norm of the propagator mismatch is bounded by the time integral of the residual divided by $\hbar$. This follows by differentiating $U_Z(t,s)J(s)U_{\rm ref}(s,0)$ and integrating. Choosing $H_Z$ to cancel the residual is not a proof that it emerged from the original source.

\section{Extended physical examples after the probability theorem}
\label{sec:I-downstream-examples}
The following examples apply the probability theorem to familiar physical preparations and readers. Each outgoing conditional state belongs to the specified instrument. The distinction between an effect and an instrument is explained in Sec.~\ref{sec:effect-instrument}.

\paragraph{Preparation verification with finite error.}
Let \(\psi\in\mathcal H_{\rm in}\) be normalized and write \(P_\psi:=|\psi\rangle\langle\psi|\).  Suppose an actual preparation is represented by a density operator \(\sigma\) satisfying
\begin{equation}
\operatorname{Tr}(\sigma P_\psi)\ge 1-\varepsilon,
\qquad 0\le\varepsilon\le1.
\end{equation}
For every effect \(0\le E\le I\), the trace-distance bound for a state compared with a pure state gives
\begin{equation}
\boxed{
\left|\operatorname{Tr}(\sigma E)-\langle\psi,E\psi\rangle\right|
\le \tfrac12\|\sigma-P_\psi\|_1
\le \sqrt{\varepsilon}.}
\label{eq:I-preparation-calibration-error}
\end{equation}
Thus imperfect verification produces a quantitative prediction error.  It does not turn a small-probability record into an impossible record, and it is not an alternative derivation of the probability law.

\paragraph{Projection, conditioning and partial trace are different operations.}
Let \(\rho\) be a density operator on a Hilbert space \(\mathcal H\) and let \(P\) be an orthogonal projector.  Then
\begin{equation}
\operatorname{Tr}_{\mathcal H}(\rho P)
=\operatorname{Tr}_{\mathcal H}(P\rho P)
=\operatorname{Tr}_{P\mathcal H}(P\rho P).
\label{eq:I-supported-trace}
\end{equation}
The last expression merely views the same supported operator on the subspace \(P\mathcal H\). It does not normalize it.  If \(p=\operatorname{Tr}(\rho P)>0\), the normalized selected state of the specific projection instrument is \(P\rho P/p\).  A partial trace is different again: it discards a specified subsystem and leaves an operator on the retained subsystem.  The instrument specifies the physical update. The effect alone specifies only its probability, as illustrated in Sec.~\ref{sec:effect-instrument}.

\paragraph{Spin readout with unresolved momentum.}
For a nonrelativistic electron write
\begin{equation}
\mathcal H_e=\mathcal H_p\otimes\mathbb C^2,
\qquad
|\Psi_e\rangle
=|\phi_+\rangle\otimes|+\rangle
+|\phi_-\rangle\otimes|-\rangle,
\label{eq:I-electron-state-example}
\end{equation}
where \(\mathcal H_p\) carries momentum wave packets and
\(\|\phi_+\|^2+\|\phi_-\|^2=1\).  The momentum vectors can be unnormalized and nonorthogonal.  Let \(I_p\) denote the identity on \(\mathcal H_p\).  An ideal sharp spin record has preparation-side effects
\begin{equation}
\begin{gathered}
\Pi_\pm=I_p\otimes|\pm\rangle\langle\pm|,\\
\Pi_++\Pi_-=I,\qquad\Pi_+\Pi_-=0.
\end{gathered}
\label{eq:I-spin-effects-example}
\end{equation}
For a source-derived detector with terminal record projections \(Q_\pm\) satisfying \(V^\dagger Q_\pm V=\Pi_\pm\), the Born theorem gives
\begin{equation}
p_\pm
=\langle\Psi_e,\Pi_\pm\Psi_e\rangle
=\|\phi_\pm\|^2.
\label{eq:I-spin-probability-example}
\end{equation}
The identity $I_p$ leaves momentum unresolved. Its physical change is specified by the instrument.

For the particular ideal projection instrument with selective map
\(\rho_e\mapsto\Pi_\pm\rho_e\Pi_\pm\), a recorded \(+\) outcome gives, for \(p_+>0\),
\begin{equation}
|\Psi_{e|+}\rangle
=\frac{\Pi_+|\Psi_e\rangle}{\sqrt{p_+}}
=\frac{|\phi_+\rangle}{\|\phi_+\|}\otimes|+\rangle.
\label{eq:I-spin-conditional-example}
\end{equation}
For this projection instrument, conditioning differs from ignoring momentum, which gives the reduced spin state
\begin{equation}
\begin{aligned}
\rho_{\rm spin}&:=\operatorname{Tr}_p\rho_e
=\sum_{r,s=\pm}\langle\phi_s|\phi_r\rangle|r\rangle\langle s|,\\
\rho_e&:=|\Psi_e\rangle\langle\Psi_e|.
\end{aligned}
\label{eq:I-spin-reduced-example}
\end{equation}
and
\begin{equation}
p_+
=\operatorname{Tr}_{\mathcal H_e}(\rho_e\Pi_+)
=\operatorname{Tr}_{\mathbb C^2}
(\rho_{\rm spin}|+\rangle\langle+|).
\label{eq:I-spin-two-traces-example}
\end{equation}
If the same projection interaction occurs but its record is not read, the corresponding nonselective state of this particular instrument is
\begin{equation}
\rho_e^{\rm unread}
=\sum_{r=\pm}\Pi_r\rho_e\Pi_r.
\label{eq:I-spin-unread-example}
\end{equation}
Decomposing a vector into components, selecting a stored record, tracing out momentum, and ignoring a detector record are therefore four different operations.

\paragraph{Hydrogen with finite energy resolution.}
For the ideal nonrelativistic Coulomb atom, with zero energy at ionization, let \(\mathcal E_H\) denote the hydrogen ionization energy and write
\begin{equation}
E_n=-\frac{\mathcal E_H}{n^2},
\qquad
\mathcal E_H\simeq13.6\,{\rm eV},
\qquad n=1,2,\ldots .
\label{eq:I-hydrogen-levels-example}
\end{equation}
The rounded constant is sufficient for the resolution example~\cite{CODATA2022}. Fine structure, radiative corrections, external fields and line widths are not included in Eq.~\eqref{eq:I-hydrogen-levels-example}.  The adjacent spacing is
\begin{equation}
E_{n+1}-E_n
=\mathcal E_H\frac{2n+1}{n^2(n+1)^2}
\sim\frac{2\mathcal E_H}{n^3}.
\label{eq:I-hydrogen-spacing-example}
\end{equation}
Hence a fixed instrumental width can eventually exceed the adjacent level spacing, without implying that all physical distinctions at high \(n\) have disappeared.

Let \(H_H\) be the internal hydrogen Hamiltonian and \(\Pi_n\) the projection onto the full bound eigenspace at \(E_n\).  An ideal energy record reporting an energy interval \(B\subset\mathbb R\) has effect
\begin{equation}
\begin{gathered}
\Pi_B:=1_B(H_H)=\sum_{E_n\in B}\Pi_n+\Pi_B^{\rm cont},\\
V^\dagger Q_BV=\Pi_B.
\end{gathered}
\label{eq:I-hydrogen-bin-example}
\end{equation}
where \(\Pi_B^{\rm cont}\) is the continuum spectral contribution when \(B\) intersects the scattering spectrum.  The probability of that single recorded energy bin is therefore
\begin{equation}
p(B)=\operatorname{Tr}(\rho\Pi_B)
=\sum_{E_n\in B}\operatorname{Tr}(\rho\Pi_n)
+\operatorname{Tr}(\rho\Pi_B^{\rm cont}).
\label{eq:I-hydrogen-probability-example}
\end{equation}
One record label may thus contain many physical states.  Coarse graining the displayed label does not identify those states, trace them out, or decide whether coherence inside the bin survives. That is fixed by the actual instrument and continuation.

As a numerical check, for \(\delta E=10^{-3}\,{\rm eV}\) the spacing \(E_{31}-E_{30}\) from Eq.~\eqref{eq:I-hydrogen-spacing-example} is approximately \(9.59\times10^{-4}\,{\rm eV}\), while the interval \([-10^{-3},0)\,{\rm eV}\) contains the Coulomb levels \(n\ge117\).  This illustrates why adjacent-level resolution and the width of a reported bin are different questions.

\paragraph{Numerical basis size is not physical resolution.}
Let \(P_N\) denote a numerical projection onto the first \(N\) basis directions of a separable represented Hilbert space.  Increasing \(N\) can improve a calculation while leaving the source-process data \((\mathcal D,\mathfrak B,s_0,\tau,\mathcal R)\) unchanged.  For spin alone two basis vectors already span \(\mathbb C^2\). A harmonic oscillator has infinitely many energy basis vectors. And including momentum gives infinitely many momentum directions inside each of the two spin effects.  None of these facts changes the number of spin records.  This is the probability version of the source reconstruction's separation between numerical refinement and physical resolution (Sec.~\ref{joint:source}).

\paragraph{Mixing and verification do not replace the statistical assumptions.}
Suppose an actual preparation register is physically included in the source.  For \(0\le t\le1\), the represented joint preparation can have the normalized state
\begin{equation}
|\Phi_t\rangle
=\sqrt t\,|\psi_0,0\rangle
+\sqrt{1-t}\,|\psi_1,1\rangle.
\label{eq:I-mixing-purification-example}
\end{equation}
Tracing over the register gives
\begin{equation}
\operatorname{Tr}_{R}|\Phi_t\rangle\langle\Phi_t|
=t|\psi_0\rangle\langle\psi_0|
+(1-t)|\psi_1\rangle\langle\psi_1|.
\label{eq:I-mixing-reduced-example}
\end{equation}
This operator identity follows from orthogonality of the physical register states. By itself it does not say that the register record occurs with probability \(t\).  Likewise calibration at the endpoints does not force a linear probability rule.  For \(x\in[0,1]\), the nonlinear function
\begin{equation}
h(x)=\frac{x^2}{x^2+(1-x)^2}
\label{eq:I-nonlinear-h-example}
\end{equation}
obeys \(h(0)=0\), \(h(1)=1\) and \(h(x)+h(1-x)=1\), yet it fails affine mixing.  For \(x_0=0\), \(x_1=1/2\), and \(t=1/2\),
\begin{equation}
h(1/4)=1/10
\ne \tfrac12 h(0)+\tfrac12 h(1/2)=1/4.
\end{equation}
The example shows why the physical mixing and preparation-sufficiency conditions used in Sec.~\ref{sec:I-born-main} are substantive statistical consistency conditions rather than consequences of state preparation alone.

An ideal verification interaction can also be written explicitly.  Let a two-state detector have a unitary flip \(X_D\) and an initial state \(|0\rangle_D\).  Then
\begin{equation}
U_{\rm test}
=P_\psi\otimes I_D+(I-P_\psi)\otimes X_D
\label{eq:I-verification-unitary-example}
\end{equation}
is unitary and leaves \(|\psi,0\rangle\) in the accepting detector sector while routing every \(|\phi,0\rangle\) with \(\phi\perp\psi\) to the orthogonal detector sector.  This is a mathematical apparatus model.  Whether the admitted source controls can realize it subject to the relevant conservation laws is a separate physical question. Neither the matrix identity nor certainty of verification supplies the Born law.

{\paragraph{Dependence on retained complementary labels.}
The grouped-output calculation in Appendix~\ref{app:comparison}
can be extended without changing the selected squared weight $a=1/2$.
Under Eq.~\eqref{eq:I-squared-weight-counterrule}, its elementary-record
probability is $1/2$, $2/3$ or $3/4$ when the complementary weight is
retained as one, two or three equal labels. This violates
the comparison condition, which requires independence from those retained data.
Reapplying the elementary-record formula to a merely coarsened report
would impose an additional context-independent event valuation.
Equality of the grouped squared length alone is not a physical
justification of that valuation.
}

\section{Causality and the law of retained detector records}
\label{app:causal-born}
This argument concerns finite-dimensional pure preparations and a fixed class of complete experiments. The source representation and detector effects are supplied independently. The following are hypotheses on the physical record law, not consequences of the kernel reconstruction.

For a fixed apparatus, procedures giving the same complete pure ray have the same record probabilities. Each run gives one retained joint record with a positive normalized law. Operations in a distant laboratory preserve the relevant local marginal probabilities. Adjoining a pure idle auxiliary system, or moving it between laboratories without an interaction, preserves those probabilities. A joint record whose represented amplitude is zero has probability zero. These statements apply to every comparison below on one declared domain.

The preparation and control domain includes the required pure states, local isometries, compensating unitaries, sharp readers and orthogonal copies of detector labels, with auxiliary systems of every finite dimension. Operations used to construct a comparison precede its terminal readings. Once a record is asserted retained, every continuation used in that comparison preserves it. This condition does not include arbitrary coherent reversal after the same record has been asserted retained. Exact controls, separation and transport are hypotheses. No construction of them from interacting FPT-QED is supplied here.

\paragraph{A fixed local reader depends only on the reduced state.}
Two pure preparations with the same reduced operator on $A$ have Schmidt decompositions with the same nonzero coefficients. After adjoining pure idle factors, their partner spaces can be made the same. A unitary on the partner maps one set of Schmidt vectors to the other, and extends on the orthogonal complement. Ray sufficiency equates the law after this unitary with that of the second preparation. No-signalling equates its $A$ marginal with the first. Idle-system invariance permits the enlargements and transport used in this comparison. Thus the fixed reader on $A$ has a function $g_A(\rho_A)$ for this preparation class. No equality of response functions on different physical carriers is inferred.

\paragraph{Equal entangled branches have equal chances.}
For $|\Phi_M\rangle=M^{-1/2}\sum_{j=1}^M|j,j\rangle$, zero-amplitude calibration permits only equal joint labels. Write their probabilities as $p_j$. Apply a permutation $\pi$ to $A$ alone. The joint support is now $(\pi(j),j)$. No-signalling preserves $B$'s probabilities $p_j$, so $A$'s probability for $k$ is $p_{\pi^{-1}(k)}$. Apply the same permutation to $B$. This restores $|\Phi_M\rangle$, while no-signalling preserves $A$'s law from the preceding step. Ray sufficiency identifies the restored law with the original one. Thus $p_k=p_{\pi^{-1}(k)}$ for every permutation. Normalization gives $p_j=1/M$. The same reasoning applies to the specified orthonormal Schmidt labels of any two such carriers.

\paragraph{Rational populations for a binary reader.}
Fix a reader on $A$ with labels $1,n$, and prepare
\[
 \sqrt{m/M}\,|1\rangle|X\rangle
 +\sqrt{(M-m)/M}\,|n\rangle|Y\rangle,
\]
where $X,Y$ are orthogonal partner states and $0<m<M$. A partner isometry with a fresh auxiliary $C$ sends
\[
\begin{aligned}
 |X\rangle&\longmapsto m^{-1/2}\sum_{k\le m}|k\rangle_P|k\rangle_C,\\
 |Y\rangle&\longmapsto (M-m)^{-1/2}\sum_{k>m}|k\rangle_P|k\rangle_C.
\end{aligned}
\]
Move $C$ to the $A$ laboratory without interacting with it. The resulting vector is maximally entangled between $AC$ and $P$ in $M$ orthogonal branches. The preceding argument fixes $P$'s label probabilities at $1/M$. Changing the distant $AC$ reader to the product reader which reads $A$ leaves this marginal unchanged. In that joint experiment, zero-amplitude calibration identifies $A=1$ with $P\le m$. Ordinary addition gives $m/M$. The local-state result and the stated invariances transfer this value to the original fixed $A$ reader. The endpoint cases follow from zero-amplitude calibration and normalization.

\paragraph{Arbitrary populations by joint-event bounds.}
Let the fixed binary reader on $B$ have reduced state $\operatorname{diag}(a,1-a)$. For rationals $l\le a\le u$ consider the normalized comparison vectors
\[
\begin{aligned}
 |\Psi_l\rangle={}&\sqrt l\,|1,1,c_1\rangle
 +\sqrt{a-l}\,|n,1,c_2\rangle\\
 &+\sqrt{1-a}\,|n,n,c_3\rangle,\\
 |\Psi_u\rangle={}&\sqrt a\,|1,1,c_1\rangle
 +\sqrt{u-a}\,|1,n,c_2\rangle\\
 &+\sqrt{1-u}\,|n,n,c_3\rangle.
\end{aligned}
\]
The first two labels belong to $A,B$, and the $c_j$ are orthogonal auxiliary states. Both vectors give the same reduced state on $B$. Their $A=1$ probabilities are $l$ and $u$ by the rational result. Zero joint amplitudes imply $\{A=1\}\subseteq\{B=1\}$ in the lower comparison and the opposite inclusion in the upper one. Positivity of each joint law and sufficiency for the same $B$ reader give $l\le g_B(a)\le u$. Rational bounds of arbitrarily small width force $g_B(a)=a$.

\paragraph{The actual detector, including degenerate sectors.}
Let a normalized complete vector have a readable detector label $D$ and a correlated label copy $D'$. Write
\[
 |\Psi\rangle=\sum_s|\Psi_s\rangle,\qquad
 |\Psi_s\rangle\in\mathcal H_{\rm rest}\otimes\operatorname{Ran}Q_s^D
                 \otimes\operatorname{Ran}Q_s^{D'},
\]
and $a_s=\|\Psi_s\|^2$. The sector vectors can be internally entangled. No product vector for a whole detector sector is assumed. Fix an event $r$ with $0<a_r<1$.

For a rational $l\le a_r$, act on the distant copy with a sector-controlled isometry which appends an auxiliary binary carrier $A$ and orthogonal tags $C$. On sector $r$ append
\[
 \sqrt{l/a_r}\,|1,c_1\rangle
 +\sqrt{1-l/a_r}\,|n,c_2\rangle.
\]
On every other sector append $|n,c_{3s}\rangle$. Orthogonality of detector sectors and tags gives $\rho_A=\operatorname{diag}(l,1-l)$. Thus $p(A=1)=l$. Zero-amplitude calibration makes $A=1$ a subevent of the actual detector event $D=r$. No-signalling leaves the detector's law unchanged, giving $p(D=r)\ge l$.

For a rational $u\ge a_r$, append $|1,c_1\rangle$ on sector $r$. On each nonzero sector $s\ne r$, append
\[
 \sqrt{t_s/a_s}\,|1,c_{2s}\rangle
 +\sqrt{1-t_s/a_s}\,|n,c_{3s}\rangle,
\]
where $t_s=a_s(u-a_r)/(1-a_r)$. These numbers satisfy $0\le t_s\le a_s$ and $\sum_{s\ne r}t_s=u-a_r$. The auxiliary reduced state is $\operatorname{diag}(u,1-u)$. Now $D=r$ is a subevent of $A=1$, so $p(D=r)\le u$. Unused zero-weight sectors can be mapped isometrically to any fresh normalized tags. If $a_r=0$ or $1$, calibration and normalization settle the result directly. Hence
\[
 p(D=r)=a_r=\|Q_r^D\Psi\|^2.
\]
For an independently represented writer $V$, this is $p(r|\psi)=\langle\psi,V^\dagger Q_r^D V\psi\rangle$. Grouping labels gives the same law for unions. An independent classical selector extends it to physical mixtures by ordinary total probability. A fundamental nonunitary source requires the corresponding independently justified output preparation and record-law conditions. Its formal dilation is not a substitute for the physical carriers assumed in this argument.

Choosing $l=\lfloor Ma_r\rfloor/M$ and $u=\lceil Ma_r\rceil/M$ gives Eq.~\eqref{eq:causal-brackets}. The proof removes a separate continuity premise by using an ideal family of exact comparisons. Finite calibration alone supplies only the comparisons actually justified.

\section{Details of the source representation and selector condition}
\label{app:representation-details}
\paragraph{Scalar classification.}
The hypotheses are those stated in Sec.~\ref{sec:II-universal-reconstruction}, including the independently supplied real algebra structure.
The proof does not assume complex scalars.  Let \(a\in\mathbb S\).  Because \(\mathbb S\) is a finite-dimensional real division algebra and is commutative, the minimal polynomial of \(a\) over \(\mathbb R\) is irreducible: a nontrivial factorization would give a product of two nonzero zero divisors after evaluation at \(a\).  Real irreducible polynomials have degree one or two.  If every element has degree one, \(\mathbb S=\mathbb R\).  Otherwise some \(a\) generates a copy \(\mathbb R[a]\simeq\mathbb C\).  Since this copy is central, \(\mathbb S\) is a finite-dimensional complex vector space.  For any \(x\in\mathbb S\), complex-linear multiplication \(L_x\) has an eigenvalue \(\lambda\in\mathbb C\).  Hence \(L_{x-\lambda}\) has a nontrivial kernel, so \(x-\lambda\) is not invertible.  The division property forces \(x=\lambda\). Therefore \(\mathbb S\simeq\mathbb C\).

The real case is excluded because its involution fixes every scalar, so its unitary scalars are only \(\{\pm1\}\). The nonconstant continuous phase in Eq.~\eqref{eq:II-continuous-source-phase} cannot take values in this discrete set. It remains to determine the involution.  A unital real-algebra involution of \(\mathbb C\) fixes \(\mathbb R\) and sends \(i\) to \(i\) or \(-i\).  If it fixes \(i\), the unitary condition is \(u(\theta)^2=1\), so the unitary scalar set is only \(\{\pm1\}\). A continuous map from the connected real line into this discrete set is constant.  Equation~\eqref{eq:II-continuous-source-phase} excludes that case.  Thus \(i^*=-i\), and the involution is complex conjugation.  The conclusion uses the reversible finite-dimensional real-algebra class specified in the main text. 
\paragraph{Tensor-product isometry.}
Use the bilinear product, factorization and density conditions of Sec.~\ref{sec:II-universal-reconstruction}.
The map on algebraic simple tensors
\begin{equation}
 U_0\bigl([s_A]\otimes[s_B]\bigr)
 :=[s_A\boxtimes s_B]
 \label{eq:II-source-tensor-map}
\end{equation}
is then well defined.  Indeed, positivity of each kernel implies by Cauchy--Schwarz that a null source vector pairs to zero with every source vector. Eq.~\eqref{eq:II-source-kernel-factorization} therefore sends a null factor to a null composite.  Moreover,
\begin{align}
 &\left\langle
 U_0([s_A]\otimes[s_B]),
 U_0([t_A]\otimes[t_B])
 \right\rangle\nonumber\\
 &\qquad=
 \langle[s_A],[t_A]\rangle\,
 \langle[s_B],[t_B]\rangle ,
 \label{eq:II-source-tensor-isometry}
\end{align}
so \(U_0\) extends uniquely to an isometry
\begin{equation}
 U_{A,B}:\mathcal H_{\Sigma_A}\otimes\mathcal H_{\Sigma_B}
 \longrightarrow
 \mathcal H_{\Sigma_A\otimes\Sigma_B}.
\end{equation}

The dense-range condition then gives Eq.~\eqref{eq:II-source-tensor-product-theorem}.

\paragraph{A fixed biased selector.}
Suppose the admitted preparation set is convex and an independent selector of a fixed probability $0<\lambda<1$ gives
\[
 q(\lambda\rho+(1-\lambda)\sigma)
 =\lambda q(\rho)+(1-\lambda)q(\sigma)
\]
for every admitted pair. Put $m=(\rho+\sigma)/2$, $u=\lambda m+(1-\lambda)\rho$ and $v=\lambda\sigma+(1-\lambda)m$. Then $\lambda u+(1-\lambda)v=m$. Two uses of the assumed equality give
\[
 q(m)=[\lambda^2+(1-\lambda)^2]q(m)
       +\lambda(1-\lambda)[q(\rho)+q(\sigma)].
\]
Cancel $\lambda(1-\lambda)>0$ to obtain midpoint affinity. Boundedness supplies full affinity as in the disk proof. A fair selector is therefore a convenient sufficient realization. Independence and a known nontrivial selector law remain statistical inputs.

\end{document}